\documentclass[aps,pre,reprint,10pt,superscriptaddress,showpacs]{revtex4-2}
\usepackage{amsfonts} 
\usepackage{amsmath}
\usepackage{amssymb}
\usepackage{graphicx}
\usepackage{subfigure}
\usepackage{color}
\usepackage{color}
\usepackage{soul}
\usepackage{cancel}
\usepackage[normalem]{ulem}
\newcommand*\xbar[1]{%
	\hbox{%
		\vbox{%
			\hrule height 0.5pt 
			\kern0.5ex
			\hbox{%
				\kern-0.1em
				\ensuremath{#1}%
				\kern-0.1em
			}%
		}%
	}%
} 
\begin{document}
	\title{Thermoacoustic internal gravity wave turbulence in the Earth's lower atmosphere	}
	\author{Sukhendu Das Adhikary}
	\email{sukhendusda@gmail.com}
	\affiliation{Department of Mathematics, Siksha Bhavana, Visva-Bharati University, Santiniketan-731 235, West Bengal, India}
	\author{Amar P. Misra}  
	\homepage{Author to whom any correspondence should be addressed}
	\email{apmisra@visva-bharati.ac.in}
	\affiliation{Department of Mathematics, Siksha Bhavana, Visva-Bharati University, Santiniketan-731 235, India}
	\begin{abstract}
		We propose, for the first time, a two-dimensional model for the nonlinear coupling of internal gravity and thermal waves in the presence of temperature-dependent density inhomogeneity due to thermal expansion and thermal feedback in stratified fluids of the Earth's lower atmosphere ($0-50$ km). Such a coupling gives rise to the evolution of thermoacoustic internal gravity waves (IGWs), which are distinctive from the known IGWs in the literature. We perform numerical simulations to study the nonlinear interactions of velocity and density perturbations associated with the IGWs and thermal fluctuations corresponding to the thermal mode. We show that solitary vortices of IGWs coupled to the thermal wave can lead to thermoacoustic turbulence. We observe the formation of large-scale velocity potential flows and small-scale structures in the density and temperature profiles. Interestingly, while the wave energy spectra exhibit  power laws: $ k_x^{-1.67}$ and $ k_z^{-2.89}$, respectively, for horizontal and vertical wave numbers, in the troposphere ($0-15$ km) with negative temperature gradient, the same in the stratosphere ($15-50$ km) with positive temperature gradient tend to relax toward $k_x^{-1.83}$-horizontal and $k_z^{-1.03}$-vertical spectra. We find that while the energy spectra in the tropospheric turbulence are consistent with the observed phenomena without temperature gradients, those in the stratosphere differ.    
	\end{abstract}
	\date{\today}
	\maketitle
	\section{Introduction} \label{sec-intro}
	Global warming is the most important cause of climate change in the Earth's atmosphere. The circulation dynamics in the atmosphere, such as thermal convection, have been studied earlier. This type of motion happens because temperature differences cause changes in air density in the presence of gravity \cite{Busse1977}. Differences in heating across a fluid create temperature variations, causing the fluid to expand more in some areas than others, resulting in density changes. These changes can cause thermal instability \cite{DJAcheson_1973}. The atmospheric fluids under gravity become stratified, and in the interior, the small-scale density and pressure fluctuations can produce internal gravity waves (IGWs) \cite{Miyoshi2008Gravity, Plougonven2014IGW}. 
These waves are typically low-frequency branches of acoustic-gravity waves (AGWs) having frequency in the range $10^{-4}~\rm{s}^{-1}<\omega<1.7\times 10^{-2}~\rm{s}^{-1}$ and wavelengths about $10$ km \cite{Kaladze2008acoustics}. These buoyancy-driven oscillations propagate in a stratified fluid, where the gravity acts as a restoring force, and have a dispersion relation similar to low-frequency ion-acoustic or ion waves in plasmas \cite{chen1983}. Even though the acoustic contribution is absent in IGWs, the term ``acoustic-gravity wave" is frequently used to encompass both the high-frequency (acoustic) and low-frequency (gravity) branches of AGWs that propagate in a compressible fluid under gravity \cite{ShaikhAGW2008,Stenflo1996}.  Stratified fluids exhibit more complex dynamics compared to homogeneous fluids. When the stratification is stable (with the \text{Brunt--V\"{a}is\"{a}l\"{a}} frequency, $\omega_g^2>0$), the fluid can support the propagation of gravity waves, such as IGWs. Stratified fluids may experience instability due to density variations between different layers of the atmosphere. The \text{Brunt--V\"{a}is\"{a}l\"{a}} frequency may become imaginary with a negative density gradient in such cases \cite{DJAcheson_1973}. If temperature variations arise from differential heating, leading to density changes due to thermal expansion, there could be competing effects between the temperature and density gradients. It has been shown that for convective flows, the thermal expansion can cause the temperature and density perturbation oscillations to couple, leading to Rayleigh-B{\'e}nard convective instability due to negative temperature gradient \cite{Kaladze2024temperature}. While these density perturbations are associated with AGWs, the temperature perturbations exhibit similar characteristics of thermal waves in plasmas \cite{misra2025,kryuchkov2018}. The thermal waves are typically damped by the influence of thermal diffusion or collisions between particles. The wave coupling reported by Kaladze \textit{et al.} \cite{Kaladze2024temperature} is analogous to the coupled thermoacoustic (thermal and acoustic-like wave coupling) mode observed experimentally in Ref. \cite{kryuchkov2018}, and a later theoretical development by Misra et al. \cite{misra2025} in complex plasmas. Thus, in Earth's atmospheric incompressible stratified fluids, due to temperature-dependent density inhomogeneity caused by thermal expansion, when thermal waves and AGWs or IGWs become coupled, they may be referred to as thermoacoustic waves in analogy to complex plasmas \cite{misra2025,kryuchkov2018} or thermoacoustic AGWs (or IGWs).    
\par 
 A small helical force affects a fluid with a stable temperature gradient, low Reynolds number, and gravity. With stable stratification, the fluid can develop a large-scale vortex instability. In the nonlinear regime, this instability become saturated and can give rise to many stationary spiral vortex structures, including a stationary helical soliton and a kink-type soliton structure \cite{tur2013non}. Some other works have also reported similar instabilities, but in rotating fluids caused by an unstable vertical temperature gradient and a small-scale force without helicity \cite{tur2013Rorate, Kopp2021}. Erdal \textit{et al.} \cite{Yigit2015983}  studied how internal waves generated in the lower atmosphere can cause vertical coupling between the atmosphere and ionosphere. They also examined how this wave-driven coupling affects sudden stratospheric warming and fluid circulation in the upper atmosphere.
	\par 
	Recently, Kaladze and Misra \cite{Kaladze2023thermal} examined how temperature and density gradients due to thermal expansion influence the stability of vertically stratified fluids in the Earth's lower atmosphere. They focused on the $0 < z < 50~\text{km}$ atmospheric layer, considering negative temperature gradients in the troposphere ($0 < z < 15~\text{km}$) and positive gradients in the stratosphere ($15 < z < 50~\text{km}$). Their analysis showed that the negative temperature gradient can cause instability in incompressible stratified fluids. They also derived the \text{Brunt--V\"{a}is\"{a}l\"{a}} frequency modified by the thermal expansion coefficient and identified its critical value that leads to IGW instability. In another work, Kaladze \textit{et al.} \cite{Kaladze2024temperature} considered the effect of density inhomogeneities caused by temperature variations and vertical temperature gradients of arbitrary sign, allowing for the expansion of atmospheric layers from $0$ to $50$ km and the consideration of arbitrary length scales of density and temperature inhomogeneities. They showed that the \text{Brunt--V\"{a}is\"{a}l\"{a}} frequency for IGWs was modified by this effect, leading to new instability conditions for IGWs. As a result, the \text{Rayleigh-B\'{e}nard} convective instability and modified instability growth rates occurred. The \text{Rayleigh-B\'{e}nard} convective instability allows for more efficient energy transfer between two atmospheric layers due to the effects of thermal expansion and the temperature gradient, resulting in a transition from laminar to turbulent flow \cite{Zhao2024turbulence, Sharifi2024spatiotemporal}. On the other hand, Shaikh \textit{et al.} \cite{ShaikhAGW2008} showed that IGWs can induce turbulent flows in atmospheric stratified fluids through the vortex motion and nonlinear interactions of high-frequency velocity field potential and low-frequency density perturbations of IGWs. In this context, Kaladze et al.  \citep{kaladze2022} reported the formation of solitary dipolar vortices in the nonlinear propagation of IGWs and their inhibition by the effects of Pedersen conductivity. Recently, Shavid et al. \cite{shavit2025} studied the turbulent spectrum of 2D IGWs in an anisotropic system, but in the context of oceanic fluid flows. However, none of the above investigations considered the coupling of IGWs with thermal waves due to thermal fluctuations. 
	We note that nonreciprocal interactions between fluid particles can induce positive thermal feedback, which potentially amplifies thermoacoustic waves.  	The latter can be excited in the atmosphere due to the coupling of sound-like waves (e.g., acoustic gravity waves, Internal waves, etc.) and thermal waves associated with temperature variations. However, these waves can undergo instability, i.e., amplification of waves can occur by positive thermal feedback of the medium on density and temperature variations \cite{misra2025}.
	\par 
	We aim to study the coupling of IGWs and thermal waves, and how the thermoacoustic instability can drive vortex motion of IGWs, leading to turbulence in tropospheric and stratospheric stratified fluids. We organize the manuscript as follows. In Sec. \ref{sec-basic}, we present the basic fluid model in the Boussinesq approximation and derive the nonlinear coupled equations for the vortex motion of IGWs and thermal waves. Section \ref{sec-linear} presents the linear theory of coupling between IGWs and thermal modes and associated growth rates of instabilities. However, we study the nonlinear evolution of thermoacoustic IGW-driven turbulence in tropospheric and stratospheric stratified fluids in Sec.  \ref{sec-nonlin}. Finally, we conclude the results in Sec. \ref{sec-conclusion}. 
	\section{The fluid model and nonlinear evolution equations}\label{sec-basic}
	In the Earth's lower atmosphere (with heights within $0<z<50$ km), the motion of incompressible stratified neutral rotating fluids under gravity [${\bf g}=(0,0,-g)$] can be described by the following momentum balance equation in the Boussinesq approximation \cite{Kaladze2024temperature}.
	\begin{equation}\label{eq-momentum1}
		\frac{\partial \mathbf{v}}{\partial t} + (\mathbf{v}\cdot\nabla)\mathbf{v}=-\frac{1}{\bar{\rho}}\nabla p+\beta T^\prime {g}\hat{z}-2{\bf \Omega}\times {\bf v},
	\end{equation}
	where ${\bf v}$, $p$, and $T^\prime({\bf r},t)~[=T({\bf r},t)-\overline{T}(z)]$ with ${\bf r}=(x,y,z)$, respectively, denote the perturbed velocity, pressure, and thermodynamic temperature of neutral fluids in which $\overline{T}(z)$ is the unperturbed temperature.  Typically, fluid density variations for IGWs do not exceed one to four per cent, i.e., the ratio of the perturbed to unperturbed density is small, $\rho^\prime/\bar{\rho}\approx 0.01-0.04$ \cite{Kaladze2023thermal}. Also, to the first-order smallness of perturbations, the equation of state (temperature-dependent density variation),  $\rho({\bf r},t)=\bar{\rho}(z)\left[1-\beta T^\prime ({\bf r},t)\right]$, where $\beta=-(1/\bar{\rho})\left(\partial \rho/\partial T\right)_p$ is the thermal expansion coefficient at constant pressure,  gives  $\rho^\prime({\bf r},t)=-\bar{\rho}(z)\beta T^\prime$. Thus, the pressure gradient force and the gravity force in the momentum balance equation \eqref{eq-momentum1}, i.e., $-\nabla p+\rho{\bf g}$ reduces in the Boussinesq approximation to   \cite{Kaladze2024temperature}   $-\nabla p+\bar{\rho}\beta T^\prime {g}\hat{z}$. The last term on the right side of Eq. \eqref{eq-momentum1} represents the Coriolis force with ${\bf \Omega}=(0,0,\Omega_0)$  denoting the angular velocity of rotating fluids.  We mention that the Boussinesq approximation can be valid for weakly nonlinear waves where the fluid density variations are much smaller than the background density (e.g., less than $1$ \% to $4$ \%) and the fluid is nearly incompressible. However, for large amplitude waves or when there are large density variations (such as in atmospheric waves propagating over significant height, or in a fluid with large temperature variations), the Boussinesq approximation may not be valid \cite{vallis2017}.     In Eq. \eqref{eq-momentum1}, we have also assumed the Reynolds number to be high enough for which the viscous effects can be neglected.  
	We note that $\bar{\rho}\equiv \bar{\rho}\left(\overline{T}(z)\right)$ is the temperature-dependent unperturbed fluid density and $\beta$ is the thermal expansion coefficient that contributes to the buoyancy force [See the last term on the right-hand side of Eq. \eqref{eq-momentum1}]. Furthermore, we have assumed that the fluid density variation due to the temperature inhomogeneity is much larger than the pressure inhomogeneity.  
	\par 
	The fluid continuity equation for the total density $\rho$ and the heat equation for the temperature $T({\bf r},t)~[=\overline{T}(z)+T^\prime({\bf r},t)]$ in the presence of heat source $q(\rho,T)$, respectively, are 
	\begin{equation}\label{eq-continuity1}
		\frac{d\rho}{dt}\equiv\frac{\partial \rho}{\partial t}+{\bf v}\cdot\nabla \rho=0,~\rm{or},~\nabla\cdot {\bf v}=0,
	\end{equation}
	\begin{equation}\label{eq-tempgrad1}
		\frac{\partial T}{\partial t}+\left(\mathbf{v}\cdot \nabla\right)T=\kappa \nabla^2 T+q(\rho, T),
	\end{equation}
	where $\kappa$ is the coefficient of thermal diffusivity, and we have used the incompressibility condition to exclude the high-frequency AGW mode from the dispersion relation and to focus on the nonlinear evolution of IGWs coupled to thermal waves \cite{Kaladze2008acoustics}.   
	\par 
	At equilibrium, we assume that the heat source is given by the Fourier law: $q=-\alpha d\overline{T}/dz$, where $\alpha$ is some constant. Thus, from Eq. \eqref{eq-tempgrad1}, the unperturbed temperature profile $\overline{T}(z)$ satisfies
	\begin{equation} \label{eq-Tbar}
		\frac{d^2 \overline{T}}{dz^2}+\frac{\alpha}{\kappa}\frac{d \overline{T}}{dz} = 0,
	\end{equation}
	which yields the following solution for $\overline{T}$.
	\begin{equation} \label{eq-Tbar-sol1}
		\overline{T}(z)-T_0=\frac{C_0\kappa}{\alpha}\left(1-e^{-\alpha z/\kappa}\right),
	\end{equation} 
	where $C_0$ is an arbitrary constant, and we have used the conditions that at $z=0$, $\overline{T}=T_0$ and $d\overline{T}/dz=C_0$. In the linear approximation, Eq. \eqref{eq-Tbar-sol1} gives 
	\begin{equation} \label{eq-Tbar-sol2}
		\overline{T}-T_0=C_0z,
	\end{equation}
	which is consistent with Ref. \citep{Kaladze2024temperature} that excludes the heat source. It follows that in the linear approximation, the equilibrium temperature profile without any heat source becomes identical with that using the Fourier law of heat source.  
	\par 
	Next, we consider the two-dimensional (2D) vortex motion (in the $xz$ plane) associated with the low-frequency IGWs that are coupled to the thermal mode. So, we assume $\partial f/ \partial y=0$ for any physical quantity $f$ and $\textbf{v}=(u,0,w)$, and introduce the vorticity variable $\zeta$ and the stream function $\psi$ [using Eq. \eqref{eq-continuity1}]  as
	\begin{equation} \label{eq-zeta}
		\begin{split}
			&\zeta=\frac{\partial u}{\partial z}-\frac{\partial w}{\partial x}, \\
			&u=-\frac{\partial \psi}{\partial z}, \: w=\frac{\partial \psi}{\partial x}.
		\end{split}
	\end{equation}
	From Eq. \eqref{eq-zeta}, the vorticity variable $\zeta$ can be rewritten as $\zeta=-\left(\partial^2/\partial x^2+\partial^2/\partial z^2\right)\psi\equiv-\nabla^2 \psi$. Next, considering the total density $\rho(x,z,t)=\bar{\rho}(z)+\rho^\prime(x,z,t)$ and total temperature $T(x,z,t)=\overline{T}(z)+T^\prime(x,z,t)$, and following Ref. \cite{Kaladze2008acoustics}, Eqs. (\ref{eq-momentum1})-(\ref{eq-tempgrad1}) reduce to
	\begin{equation} \label{eq-momentum2}
		\begin{split}
			&\bar{\rho} \left[ \frac{\partial}{\partial t} \nabla^2 \psi +J(\psi, \nabla^2 \psi) \right]\\ &=\bar{\rho} \beta g \frac{\partial {T^\prime}}{\partial x}-\frac{d \bar{\rho}}{dz} \left[ \frac{\partial}{\partial t} \left( \frac{\partial \psi}{\partial z} \right)+ J \left( \psi, \frac{\partial \psi}{\partial z}\right) \right],
		\end{split}
	\end{equation}
	\begin{equation} \label{eq-continuity2}
		\frac{\partial {\rho^\prime}}{\partial t}+\frac{\partial \psi}{\partial x} \frac{d \bar{\rho}}{dz}+J(\psi,  \rho^\prime)=0,
	\end{equation}
	\begin{equation} \label{eq-tempgrad2}
		\begin{split}
			\frac{\partial {T^\prime}}{\partial t}
			+ \frac{\partial \psi}{\partial x} \frac{d \overline{T}}{dz}
			+ J(\psi, {T^\prime})
			&= \kappa \left[ \nabla^2 {T^\prime} + \frac{\partial^2 \overline{T}}{\partial z^2} \right] \\
			&\quad + q_T {T^\prime} + q_{\rho} {\rho^\prime},
		\end{split}
	\end{equation}
	where $J(a,b)=\left(\partial a/\partial x\right)\left(\partial b/\partial z\right)-\left(\partial b/\partial x\right)\left(\partial a/\partial z\right)$ denotes the Jacobian of $a$ and $b$, and the parameters $q_{T_0}=(\partial q/\partial T)_0$ and $q_{\rho_0}=(\partial q / \partial \rho)_0$, calculated at the equilibrium values $\overline{T}$ and $\bar{\rho}$, respectively, represent the thermal feedback of the medium on the temperature and density perturbations. By Taylor expanding $q(\rho,T)$ about $\left(\bar{\rho},\overline{T}\right)$ and retaining terms up to the first-orders of magnitudes, we have $q(\rho,T)=q_{\rho_0} {\rho^\prime}+q_{T_0} {T^\prime}$. From Eqs. \eqref{eq-momentum2}-\eqref{eq-tempgrad2}, we note that the contribution from the Coriolis force disappears due to the choice of the angular velocity ${\bf \Omega}$ along the $z$-axis and the fluid motion in the $xz$-plane.  
	\par
	From Eq. \eqref{eq-momentum1}, it can be assessed that when the equilibrium density profile  $\bar{\rho}$ exponentially decays, the amplitude profiles of the velocity components exponentially increase but the temperature profile decreases.
	So, similar to Ref. \citep{Kaladze2008acoustics}, we assume  the solutions of Eqs. (\ref{eq-momentum2})-(\ref{eq-tempgrad2}) in the following form: 
	\begin{equation} \label{eq-sol-form}
		\psi= e^{z/2H} \widetilde{\psi}, ~ {\rho^\prime}= e^{-z/2H} \widetilde{\rho}, ~ {T^\prime}= e^{-z/2H} \widetilde{T},
	\end{equation}
	where $H$ is the vertical scale height to be defined later [See Eq. \eqref{eq-rho-bar1}].  
	Substituting Eq. \eqref{eq-sol-form} into Eqs. (\ref{eq-momentum2})-(\ref{eq-tempgrad2}), we obtain
	\begin{eqnarray} \label{eq-momentum3}
		& &\frac{\partial}{\partial t} \left[ \nabla^2 \widetilde{\psi} +\left(\frac{1}{H}+\frac{1}{\bar{\rho}} \frac{d \bar{\rho}}{d z} \right)\frac{\partial \widetilde{\psi}}{\partial z}+\left(\frac{1}{4H^2}+\frac{1}{2 H \bar{\rho}} \frac{d \bar{\rho}}{d z}\right) \widetilde{\psi} \right] \nonumber \\
		& &+e^{z/2H}J\left(\widetilde{\psi},\nabla^2 \widetilde{\psi}\right)+e^{z/2H}\left(\frac{1}{H}+\frac{1}{\bar{\rho}} \frac{d \bar{\rho}}{d z} \right)J \left(\widetilde{\psi},\frac{\partial \widetilde{\psi}}{\partial z}\right) \nonumber \\
		& &=-e^{-z/H} \beta g\frac{\partial \widetilde{T}}{\partial x},
	\end{eqnarray}
	\begin{equation} \label{eq-continuity3}
		\frac{\partial \widetilde{\rho}}{\partial t}+J\left(\widetilde{\psi},\widetilde{\rho}\right)=-e^{z/H} \frac{d \bar{\rho}}{dz} \frac{\partial \widetilde{\psi}}{\partial x},
	\end{equation}
	\begin{equation} \label{eq-tempgrad3}
		\begin{split}
			\frac{\partial \widetilde{T}}{\partial t}+&J\left(\widetilde{\psi},\overline{T}\right)=-e^{z/H} \frac{d \overline{T}}{dz} \frac{\partial \widetilde{\psi}}{\partial x}+\kappa e^{z/2H} \frac{d^2 \overline{T}}{d z^2}\\
			&+\kappa \nabla^2 \widetilde{T}-\frac{\kappa}{H}\frac{\partial \widetilde{T}}{\partial z}+\left(\frac{\kappa}{4 H^2}+q_{T_0}\right)\widetilde{T}+q_{\rho_0}\widetilde{\rho}.
		\end{split}
	\end{equation}
	\par 
	Typically, the thermal fluctuation in the Earth's atmosphere causes density variation due to thermal expansion. So, we model the unperturbed density $\bar{\rho}(z)$ as 
	\begin{equation} \label{eq-rho-bar}
		\bar{\rho}(z)=\rho_0\exp\left[-\beta\left(\overline{T}(z)-T_0\right)\right],
	\end{equation}
	where $\rho_0$ and $T_0$ are the reference density and temperature such that $\bar{\rho}=\rho_0$ at $\overline{T}=T_0$. In the linear approximation, Eq. \eqref{eq-rho-bar} agrees with those considered in Refs. \citep{DJAcheson_1973,Kaladze2024temperature}. 
	From Eq. \eqref{eq-Tbar-sol2}, $C_0=d\overline{T}/dz$. Also, $C_0<0~(>0)$ in the troposphere (stratosphere) and $d\bar{\rho}/dz<0$ in both the tropospheric and stratospheric regions \citep{Kaladze2023thermal}. So, we choose the arbitrary constant $C_0$ such that $\beta |C_0|\sim1/H$, and Eq. \eqref{eq-rho-bar} reduces to \cite{Kaladze2008acoustics}
	\begin{equation} \label{eq-rho-bar1}
		\bar{\rho}(z)=\rho_0e^{-\beta |C_0|z}\sim\rho_0e^{-z/H}.
	\end{equation}
	Typically, IGWs are buoyancy-driven waves that propagate in the atmospheric stratified fluids and due to density stratification their vertical motion is restricted, implying that the vertical wavelength of IGWs can be much smaller than the horizontal wavelength, i.e., $k_z\gg k_x$, or equivalently,  $k_z^2\gg1/4H^2$ (Boussinesq approximation) \cite{vallis2017}.  Thus, similar to Ref. \citep{Kaladze2008acoustics}, we assume that $\exp\left(z/2H\right)\approx1$ for the nonlinear terms  and note $\beta e^{-z/H}\sim \left(1-z/H+\cdots\right)/|C_0|H\sim1/|C_0|H\sim\beta$, $d^2\overline{T}/dz^2\approx0$ (consistent with the linear approximation of $\overline{T}$), $e^{z/H}\left(d\bar{\rho}/dz\right)\sim-\rho_0/H$,  $ e^{z/H}\left(d\overline{T}/dz\right)\sim C_0e^{z/H}\sim \left(1+z/H+\cdots\right)/\beta H\sim \pm 1/\beta H\sim C_0$.  
	\par 
	Next, introducing the new variable $\widetilde{\chi}=g\widetilde{\rho}/\rho_{0}$, we obtain from Eqs. \eqref{eq-momentum3}-\eqref{eq-tempgrad3} the following reduced equations.
	\begin{eqnarray} \label{eq-mom4}
		\frac{\partial}{\partial t} \left( \nabla^2 \widetilde{\psi} -\frac{1}{4H^2}\widetilde{\psi} \right)+J(\widetilde{\psi},\nabla^2 \widetilde{\psi})=\beta g \frac{\partial \widetilde{T}}{\partial x},
	\end{eqnarray}
	\begin{equation} \label{eq-cont4}
		\frac{\partial\widetilde{\chi}}{\partial t}+J(\widetilde{\psi},\widetilde{\chi})=\omega_{g}^2\frac{\partial \widetilde{\psi}}{\partial x},
	\end{equation}
	\begin{equation} \label{eq-temp4}
		\begin{split}
			\frac{\partial \widetilde{T}}{\partial t}
			&+J(\widetilde{\psi},\widetilde{T})=\kappa \nabla^2 \widetilde{T}-\frac{\kappa}{H}\frac{\partial \widetilde{T}}{\partial z}\\
			&+\left(\frac{\kappa}{4 H^2}+q_{T_0}\right)\widetilde{T}+\alpha_{0}\widetilde{\chi}
			-C_0 \frac{\partial \widetilde{\psi}}{\partial x}, 
		\end{split}
	\end{equation}
	where $\alpha_{0}=\rho_0q_{\rho_0}/g$ and $\omega^2_g\equiv-\left[1/\bar{\rho}(z)\right]\left[d\bar{\rho}(z)/dz\right]g\approx{g/H}$ [using Eq. \eqref{eq-rho-bar1}] is the squared Brunt--V\"{a}is\"{a}l\"{a} frequency.  
	Before we study the system of Eqs. \eqref{eq-mom4}--\eqref{eq-temp4} for the coupling between IGWs and thermal modes and thermoacoustic instability, as well as, the nonlinear evolution of thermoacoustic IGWs, it is pertinent to recast the system in a dimensionless form. So, we redefine the variables and some parameters as
	\begin{equation}\label{eq-normal}
		\begin{split}
			(x,z) &\rightarrow (x,z)/H, \quad t \rightarrow t\omega_{g}, \\
			\widetilde{\psi} \rightarrow \widetilde{\psi}&/\psi_{0}, \quad \widetilde{\chi} \rightarrow \widetilde{\chi}/\chi_{0}, \quad \widetilde{T} \rightarrow \widetilde{T}/T_0, \\
			\alpha^\prime &= \frac{\psi_{0}}{\omega_{g} H^2}, \quad \zeta^\prime = \frac{\omega_{g} \psi_{0}}{H \chi_{0}}, \\
			\beta^\prime &= \frac{\beta T_0 g}{\omega_{g}^2 H \alpha^\prime}, \quad \kappa^\prime = \frac{\kappa}{\omega_{g} H^2}.
		\end{split}
	\end{equation}
	Thus, with the normalization in Eq. \eqref{eq-normal}, Eqs. \eqref{eq-mom4}--\eqref{eq-temp4} reduce to
	\begin{equation} \label{eq-mom5}
		\frac{\partial}{\partial t} \left( \nabla^2 \widetilde{\psi} -\frac{1}{4}\widetilde{\psi} \right)+\alpha^\prime J\left(\widetilde{\psi},\nabla^2 \widetilde{\psi}\right)=\beta^\prime \frac{\partial \widetilde{T}}{\partial x},
	\end{equation}
	\begin{equation}\label{eq-cont5}
		\frac{\partial \widetilde{\chi}}{\partial t}+\alpha^\prime J\left(\widetilde{\psi},\widetilde{\chi}\right)=\zeta^\prime\frac{\partial \widetilde{\psi}}{\partial x},
	\end{equation}
	\begin{equation} \label{eq-temp5}
		\begin{split}
			\frac{\partial \widetilde{T}}{\partial t}+\alpha^\prime J\left(\widetilde{\psi},\widetilde{T}\right)=&\kappa^\prime \left(\nabla^2 \widetilde{T}-\frac{\partial \widetilde{T}}{\partial z}\right)+A_1\widetilde{T}\\
			&+A_2\widetilde{\chi}-A_3 \frac{\partial \widetilde{\psi}}{\partial x},
		\end{split}
	\end{equation}
	where $A_1=\left(\kappa^\prime/4+q^\prime_{T_0}\right)$, $q^\prime_{T_0}=q_{T_0}/\omega_g$,  $A_2=\left(\alpha_{0}\chi_{0}/{\omega_{g}T_{0}}\right)$, and $A_3=\alpha^\prime C_0/T_0$.
	Equations \eqref{eq-mom5}--\eqref{eq-temp5} are the desired set of coupled equations describing the nonlinear evolution of solitary vortices in the coupling of high-frequency velocity potential field and low-frequency density fluctuations of IGWs with the thermal mode. Rewriting the temperature perturbation in terms of the density perturbation, i.e., $T^\prime=-\rho^\prime/\beta\bar{\rho}$ and ignoring the thermal fluctuation and heat equation [Eq. \eqref{eq-tempgrad1}], one can recover the known result of IGW-driven solitary vortices \cite{Stenflo1996}. These solitary vortices are typically localized vortex structures whose formation and propagation are sustained by internal gravity waves in a stratified fluid medium (See, e.g., Refs. \cite{Kaladze2008acoustics,kaladze2022}).  Thus, Eqs. \eqref{eq-mom5}--\eqref{eq-temp5} significantly advance the previous theory of IGWs with the effects of thermal gradients and thermal feedback of the medium on density and temperature fluctuations.
	In Secs. \ref{sec-linear} and \ref{sec-nonlin}, we will study the linear theory of thermoacoustic IGWs and nonlinear evolution of IGW-driven thermoacoustic turbulence by simulation approach.
	\section{Linear regime: Mode coupling and instability} \label{sec-linear}
	\par 
	In the linear regime, assuming the physical quantities to vary as plane waves with constant amplitudes in the form $\sim\exp[i(k_x x+k_z z-\omega t)]$, where $\omega$ is the wave frequency and $k=\sqrt{k_x^2+k_z^2}$ is the wave number, we obtain from Eqs. \eqref{eq-mom5}--\eqref{eq-temp5}, the following dispersion relation.
	\begin{equation}\label{eq-dr1}
		\begin{split}
			&\left[\omega^2\left(k^2+{1}/{4}\right)-A_3\beta^\prime k_x^2 \right]\left[\omega-k_z\kappa^\prime+i\kappa_q \right]\\&+i\beta^\prime k_x^2\left[A_3\left(\kappa_q+i\kappa^\prime k_z \right)+A_2\zeta^\prime\right]=0,
		\end{split}
	\end{equation} 
	where $\kappa_q=\kappa^\prime(k^2-1/4)-q_{T_0}$. The first factor of Eq. \eqref{eq-dr1} represents the IGWs, while the second factor corresponds to the thermal mode.  We note that the IGWs exist and couple with the thermal wave by the influence of the parameter $\beta^\prime$. It follows that thermal expansion plays a key role in the coupling of IGWs and thermal waves, especially in situations where significant thermal gradients and density stratification occur in the Earth's atmosphere. We also note that the thermal wave frequency gets down-shifted (by $k_z\kappa^\prime$) due to the thermal diffusivity. In the absence of the latter and the thermal feedback, we can recover only the IGW mode in stable stratified stratosphere $(C_0>0)$ \citep{Kaladze2024temperature} with the frequency, given by,
		\begin{equation} \label{eq-disp-IGW}
		\omega^2= \Omega_{\rm{IG}}^2\equiv\frac{\omega_g^2 k_x^2}{k^2+{1}/{4}}\approx  \frac{\omega_g^2 k_x^2}{k^2},
	\end{equation}
	where we have used the relation $A_3\beta^\prime=\omega_g^2$ and the  Boussinesq approximation, $k\gg1/2$ \cite{vallis2017}. 
	Equation \eqref{eq-disp-IGW} is consistent with the known result for low-frequency IGWs in stable stratified atmospheric fluids  \citep{Kaladze2008acoustics}. However, in contrast to Ref. \citep{Kaladze2008acoustics}, the mode in Eq. \eqref{eq-disp-IGW} appears due to the temperature dependent density inhomogeneity associated with thermal expansion. Equation \eqref{eq-dr1} shows that the dispersion properties of IGWs get significantly modified by the influence of the thermal wave. They become unstable by the effects of the thermal diffusivity and thermal feedback. We mention that IGWs are sometimes called AGWs in general \cite{ShaikhAGW2008,Stenflo1996}. The second factor in the first term on the left side of Eq. \eqref{eq-dr1}, when set to zero, gives a dispersion relation similar to a thermal wave in complex plasmas \cite{misra2025}. We call the coupled modes (IGWs and thermal waves) as thermoacoustic internal gravity waves.  
	\par 
	To study the dispersion properties and growth rates of instabilities of thermoacoustic IGWs (i.e., IGWs modified by the thermal mode), we assume that the coupling term $(\propto\beta^\prime)$ in Eq. \eqref{eq-dr1} is small. Thus, we obtain from Eq. \eqref{eq-dr1} the following approximate expressions for the real $(\omega_r)$ and imaginary parts $(\gamma)$ of the wave frequency $\omega$ of thermoacoustic IGWs in tropospheric ($C_0<0$) and stratospheric ($C_0>0$) fluids.
	\begin{equation} \label{eq-dr2}
		\begin{split}
			&\omega_r= \frac{1}{\sqrt{2}}\left(\frac{\beta^\prime k_x^2}{k^2+1/4}\right)^{1/2}\\
			&\times\left[\left(A_3-\frac{\lambda_1}{\lambda^2}\right)+\sqrt{\left(A_3-\frac{\lambda_1}{\lambda^2}\right)^2+\frac{\lambda_2^2}{\lambda^4}} \right]^{1/2},\\
			&\gamma=-\frac{1}{2}\frac{\beta^\prime k_x^2 \lambda_2}{\left(k^2+1/4\right)\omega_r\lambda^2},
		\end{split}
	\end{equation}
	where the expressions for $\lambda_1$, $\lambda_2$, and $\lambda$ are given by
	\begin{equation}
		\lambda_1=\left\lbrace 
		\begin{array}{cc}
			\left(A_3\kappa_q+A_2\zeta^\prime\right)\kappa_q-\left(\Omega_{\rm{IG}}-k_z\kappa^\prime\right)A_3\kappa^\prime k_z,  & C_0>0, \\ 
			A_3\left(\kappa^\prime k_z\right)^2+\left(A_3\kappa_q+A_2\zeta^\prime\right)\left(\kappa_q+\Omega_{\rm{IG}}\right), & C_0<0, 
		\end{array} \right.
	\end{equation}
	\begin{equation}
		\lambda_2=\left\lbrace 
		\begin{array}{cc}
			\left(A_3\kappa_q+A_2\zeta^\prime\right)\left(\Omega_{\rm{IG}}-k_z\kappa^\prime\right)+\kappa_q A_3\kappa^\prime k_z,  & C_0>0, \\ 
			A_3\kappa^\prime k_z\left(\kappa_q+\Omega_{\rm{IG}}\right)-\kappa^\prime k_z\left(A_3\kappa_q+A_2\zeta^\prime\right), & C_0<0, 
		\end{array} \right.
	\end{equation}
	\begin{equation}
		\lambda^2=\left\lbrace 
		\begin{array}{cc}
			\left(\Omega_{\rm{IG}}-k_z\kappa^\prime\right)^2+\kappa_q^2,  & C_0>0, \\
			\left(\kappa^\prime k_z\right)^2+\left(\kappa_q+\Omega_{\rm{IG}}\right)^2,  & C_0<0.
		\end{array} \right.
	\end{equation}
	\par
\begin{figure*}
		\includegraphics[scale=0.4]{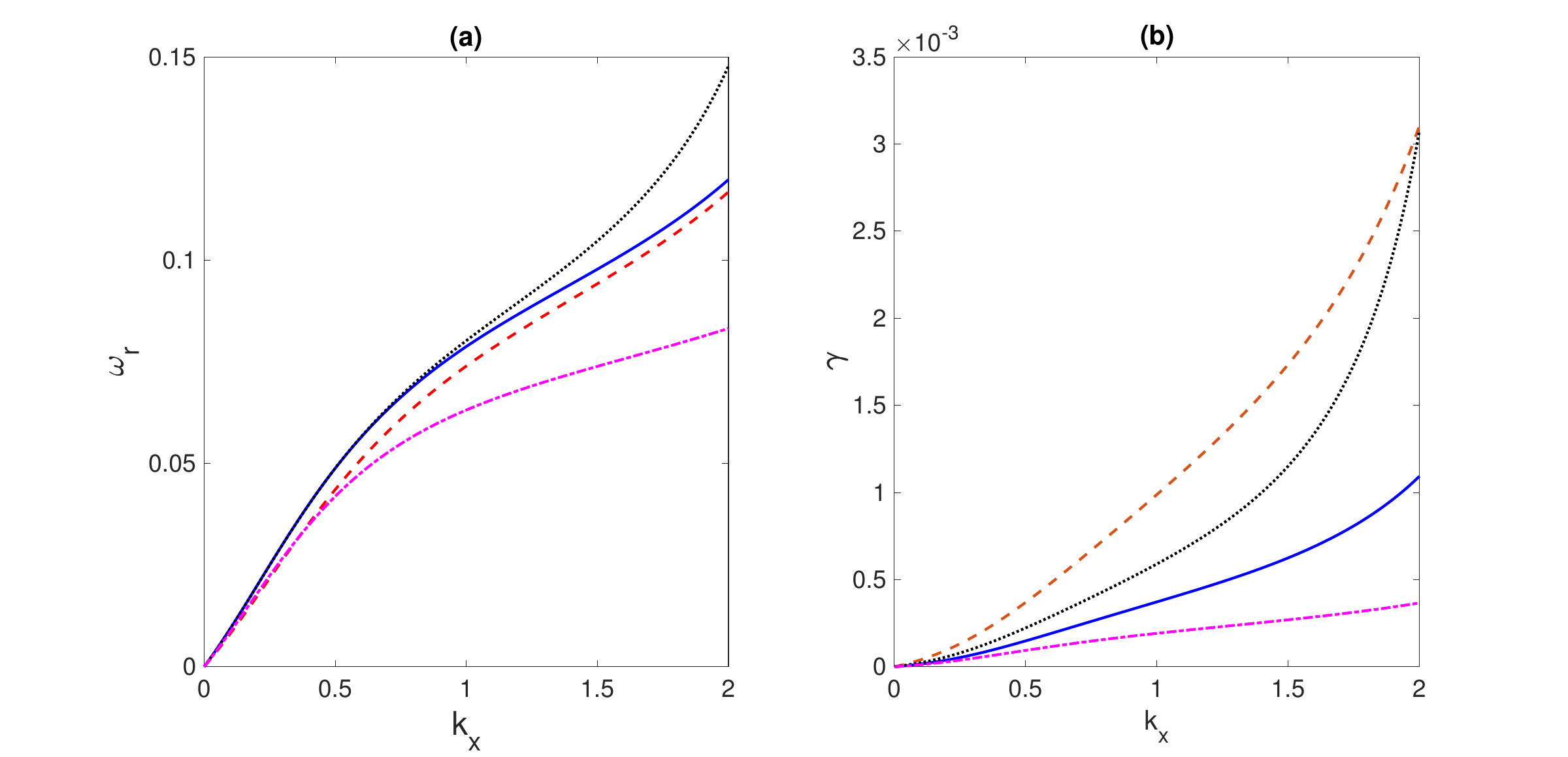}
		\caption{Dispersion curves for the real wave frequency $(\omega_r)$ and the instability growth rate $(\gamma)$ of thermoacoustic internal gravity waves in the case of $C_0\equiv d\overline{T}/dz<0$ are shown for different parameter values. The solid, dashed, dotted, and dash-dotted lines, respectively, correspond to $k_z=0.1,~\kappa^\prime=0.00085,~q_{T_0}=0.03$; $k_z=0.3,~\kappa^\prime=0.00085,~q_{T_0}=0.03$; $k_z=0.1,~\kappa^\prime=0.0013,~q_{T_0}=0.03$; and $k_z=0.1,~\kappa^\prime=0.00085,~q_{T_0}=0.035$.  The other fixed parameter values are as in Table \ref{tab-parameter} for $C_0<0$. } \label{fig1-disp}
	\end{figure*}

\begin{figure*} 
		\includegraphics[scale=0.4]{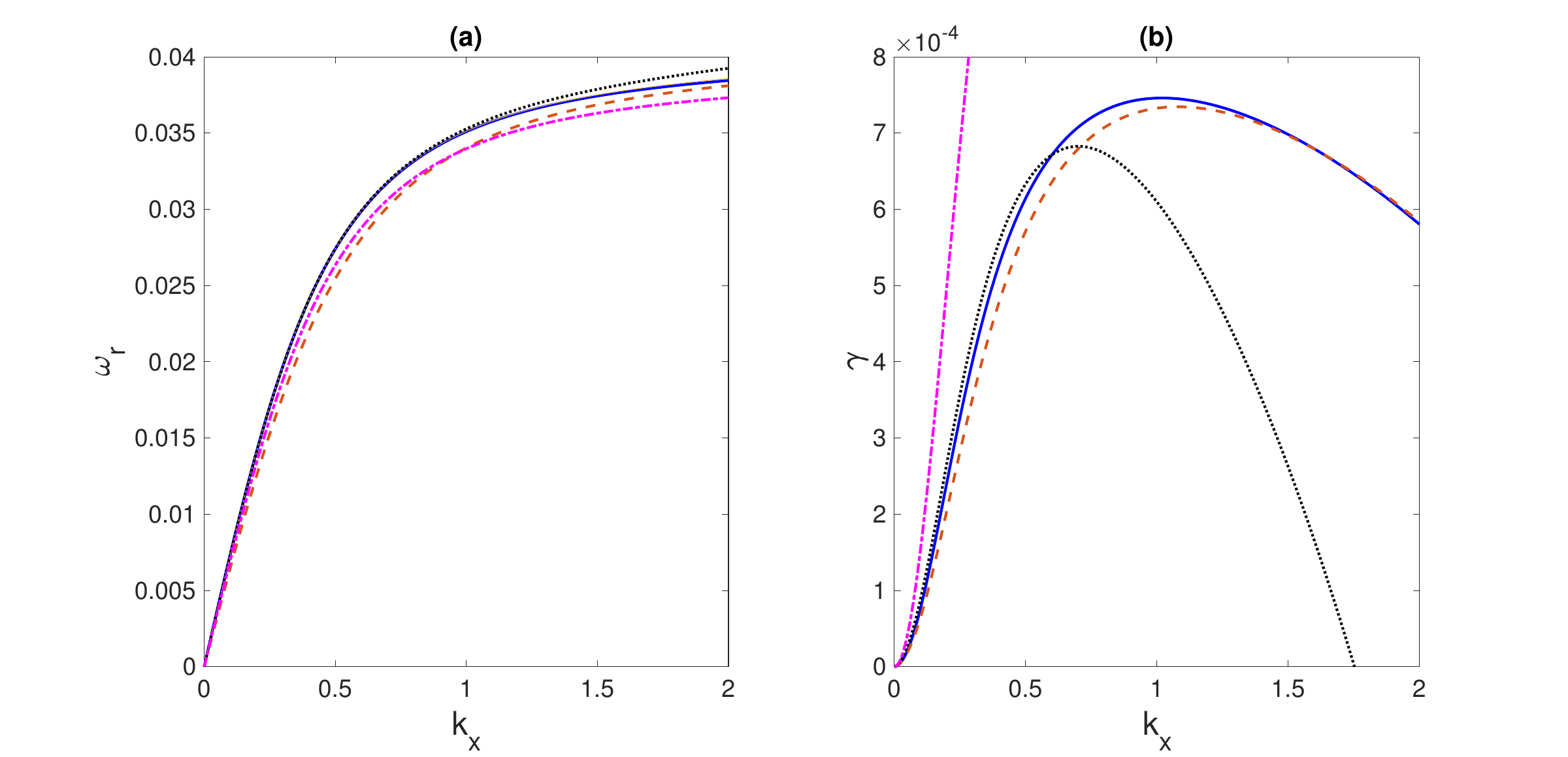}
		\caption{Dispersion curves for the real wave frequency $(\omega_r)$ and the instability growth rate $(\gamma)$ of thermoacoustic internal gravity waves in the case of $C_0\equiv d\overline{T}/dz>0$ are shown for different parameter values. The solid, dashed, dotted, and dash-dotted lines, respectively, correspond to $k_z=0.1,~\kappa^\prime=0.000279,~q_{T_0}=0.0355$; $k_z=0.3,~\kappa^\prime=0.000279,~q_{T_0}=0.0355$; $k_z=0.1,~\kappa^\prime=0.001,~q_{T_0}=0.0355$, and $k_z=0.1,~\kappa^\prime=0.000279,~q_{T_0}=0.04$.  The other fixed parameter values are as in Table \ref{tab-parameter} for $C_0>0$.} \label{fig2-disp}
	\end{figure*}

 
	In what follows, we numerically study the dispersion properties and instability growth rates of thermoacoustic IGWs and exhibit their profiles in Figs. \ref{fig1-disp} and \ref{fig2-disp} for $C_0<0$ and $C_0>0$ respectively. We note that in both cases, the low-frequency thermoacoustic IGWs propagate with distinct significant features. In the case of $C_0<0$ (Fig. \ref{fig1-disp}), which is typically the case of unstable stratified fluids in the troposphere \citep{Kaladze2024temperature}, both the real wave frequency and the instability growth rate increase with the wave number $k_x$. However, increasing the thermal feedback can result in a reduction of both the frequency and the growth rate (See the dash-dotted lines) of IGWs. Physically, since IGWs are buoyancy-driven waves where the buoyancy provides the restoring force, as the thermal feedback increases, the buoyancy frequency decreases, resulting in a decay of the wave frequency. Similarly, stronger thermal feedback can reduce atmospheric stability in addition to increasing the dissipation of wave energy, which, in turn, inhibits the growth rate of instability. We also observe a reduction of the wave frequency but a higher growth rate with increasing values of the vertical wavenumber $k_z$ (See the dashed lines). Typically, the longer the vertical wave number, the shorter the vertical wavelength, and IGWs with shorter wavelengths can become unstable more easily, leading to an increased instability growth rate. It is interesting to note that although thermal diffusivity plays a role in wave damping \citep{misra2025}, in the presence of a negative thermal gradient, the net effect is that it manifests an increase of both the wave frequency and the growth rate through its impact on thermal diffusion and wave dissipation (See the dotted lines). However, in Fig. \ref{fig2-disp}, we will see that the thermal diffusivity indeed plays a role of wave dissipation in the presence of a positive thermal gradient. 
	\par 
	Figure \ref{fig2-disp} shows that the typical acoustic-like features (e.g., ion-acoustic waves in plasmas \cite{chen1983}) of IGWs (i.e., the wave frequency approaches a constant value with increasing values of the wave number) remain preserved in the case of $C_0>0$ even with the thermal effects. The presence of a heat source and thermal feedback causes instability; however, the instability growth rate can have a cut-off at a finite wave number due to thermal diffusivity effects (See the dotted lines). Physically, higher thermal diffusivity results in a faster rate of heat transfer, which can dampen the instability associated with thermal fluctuations and become significant in certain situations, especially when the thermal diffusivity is high enough to restrain the thermal fluctuations. However, the wave frequency can get reduced and the instability growth rate can be enhanced without any cut-off within a finite domain of $k_x$ by the effects of an increased thermal feedback (See the dash-dotted lines). We also note that in contrast to the case of $C_0<0$, the influence of an increased vertical wave number is to decrease both the wave frequency and the growth rate (See the dashed lines).  
	\par 
	Thus, the motion of low-frequency thermoacoustic IGWs in both the cases of negative and positive thermal gradients is unstable, and the instability growth rates remain high for long-wavelength perturbations. However, the growth rate can have a cut-off only in the case of $C_0>0$.
	\section{Nonlinear evolution of turbulence: Simulation approach}\label{sec-nonlin}
	\par Equations \eqref{eq-mom5}-\eqref{eq-temp5} admit the following energy integral \cite{Kaladze2008acoustics}.
	\begin{equation}\label{eq-energy}
		E=\int{\left[(\nabla\widetilde{\psi})^2+\frac{1}{4}\widetilde{\psi}^2+\widetilde{\chi}^2+\frac{\beta^\prime}{A_3}\widetilde{T}^2\right]}dxdz,
	\end{equation}
	where the integral proportional to $\beta^\prime$ can be negative or positive according to when $C_0=d\overline{T}/dz<0$ (in the troposphere) or $C_0>0$ (in the stratosphere). 
	We note that with the integrals corresponding to the vorticity–stream function $\widetilde{\psi}$ and the density fluctuation $\widetilde{\chi}$, an additional energy integral appears due to the temperature variation $\widetilde{T}$. In the presence of the latter, the system energy transfer rate can be faster and can reach a steady state over a longer time. To ensure the numerical validity of our code, we monitor the energy integral \eqref{eq-energy}, which should remain conserved under ideal conditions, i.e., without any source of energy dissipation \cite{shaikh2009_simulation}. However, in the present model, the IGWs interact with the thermal waves, which can result in the transfer of wave energy faster due to instability of thermoacoustic IGWs. 
	\par 
	In what follows, we study the nonlinear interaction between IGWs and thermal waves associated with temperature fluctuations through a numerical simulation approach, considering the temperature-dependent density inhomogeneity due to thermal expansion and thermal feedback. These interactions not only influence the propagation characteristics of both waves but are significant in understanding the energy transfer between them and between high- and low-frequency parts of velocity and density spectra, as well as turbulent mixing in atmospheric stratified fluids, leading to complex energy cascades.   Previous studies have focused either on the evolution of acoustic vortices \cite{kaladze2022} or the emergence of IGW turbulence  \citep{ShaikhAGW2008} without the effects of coupling between IGWs and the thermal wave or thermal expansion. We aim to study the vortex turbulent spectra of IGWs coupled to thermal waves, i.e., thermoacoustic IGWs, which are governed by Eqs. \eqref{eq-mom5}-\eqref{eq-temp5} and are relevant in the troposphere and stratosphere.   
	\par
	To investigate the nonlinear interactions of IGWs and thermal waves and the evolution of thermoacoustic IGW turbulence, we have used a spectral method \cite{Shaikh_2006, ShaikhAGW2008} to numerically integrate and solve the normalized equations \eqref{eq-mom5}-\eqref{eq-temp5}. In our code, we employ a discrete Fourier representation of the turbulent spectra to discretize the spatial domain and choose the initial isotropic turbulent spectrum in random phases in all directions. The initial fluctuations do not involve any flows or mean fields. We advance the equations in time using the Runge-Kutta fourth-order scheme with the time step $dt=0.001$ and spatial grid size $512\times512$ and make the code stable by properly using the dealiasing techniques to the spurious Fourier modes \cite{ShaikhAGW2008}. We take the time and length scale as $\omega_g\sim0.05~\rm{s}^{-1}$, $H\sim4\times 10^3$ m, respectively, also employ the energy conservation principle to verify the numerical accuracy and validity of the code during the nonlinear evolution of wave turbulence. We note that the temperature variation in the Earth's atmosphere can result in the positive and negative thermal gradients \cite{Kaladze2023thermal}, i.e., $C_0\equiv d\overline {T}/dz<0$ in the tropospheric height $0<z<15$ km and $C_0>0$ in the stratospheric region $15<z<50$ km. Accordingly, we choose the parameter values relevant to the lower atmosphere ($0 < z < 50$ km), as listed in Table \ref{tab-parameter}. Furthermore, we assume $\psi_0 \sim \omega_g H^2$ and $\chi_0 \sim \omega_g^2 H$ \cite{kaladze2022}, which give $\alpha^\prime \lesssim 1$ and $\zeta^\prime \lesssim 1$.  In Secs. \ref{sec-tropo} and \ref{sec-strato}, we investigate the turbulent spectra in these two regions separately. In each of the cases, we consider the parameter values for which the Froude number, $F_r=V/\sqrt{gH}\sim\alpha^\prime$ remains smaller than unity (\textit{cf}. Table \ref{tab-parameter}), where $V$ is the fluid velocity scale. Here, we note that if the scale height $H$ is chosen higher than that we have considered, the values of $F_r$ can be shown to be further reduced as $\alpha^\prime\sim1/H^2$.    
	\subsection{Turbulence spectra in the troposphere} \label{sec-tropo}
	We consider the evolution of thermoacoustic IGW turbulence in the tropospheric region with heights $0 < z < 15$ km. In this region, the equilibrium temperature varies with the atmospheric height $z$ almost linearly such that $C_0=d\overline{T}/dz<0$, and the equilibrium density has approximately a linear relationship with the equilibrium temperature on account of the thermal expansion \cite{Kaladze2023thermal}. It has been shown by Kaladze et al. \citep{Kaladze2024temperature} that due to this thermal expansion effect, tropospheric stratified fluids can undergo instability. In Sec. \ref{sec-basic}, we have also seen that the IGW, coupled to the thermal mode, can be unstable due to the influences of the negative temperature gradient and thermal feedback. Thus, in nonlinear interactions, the energy transport between high- and low-frequency perturbations associated with velocity and density perturbations, as well as between the IGW and thermal mode, can lead to the growth of these unstable modes and eventually turbulence. The energy transfer from larger to smaller scales will become faster as the instability develops more rapidly in the system.
	\par
\begin{figure*}[!ht]
		\includegraphics[width=6.5in, height=2.12in]{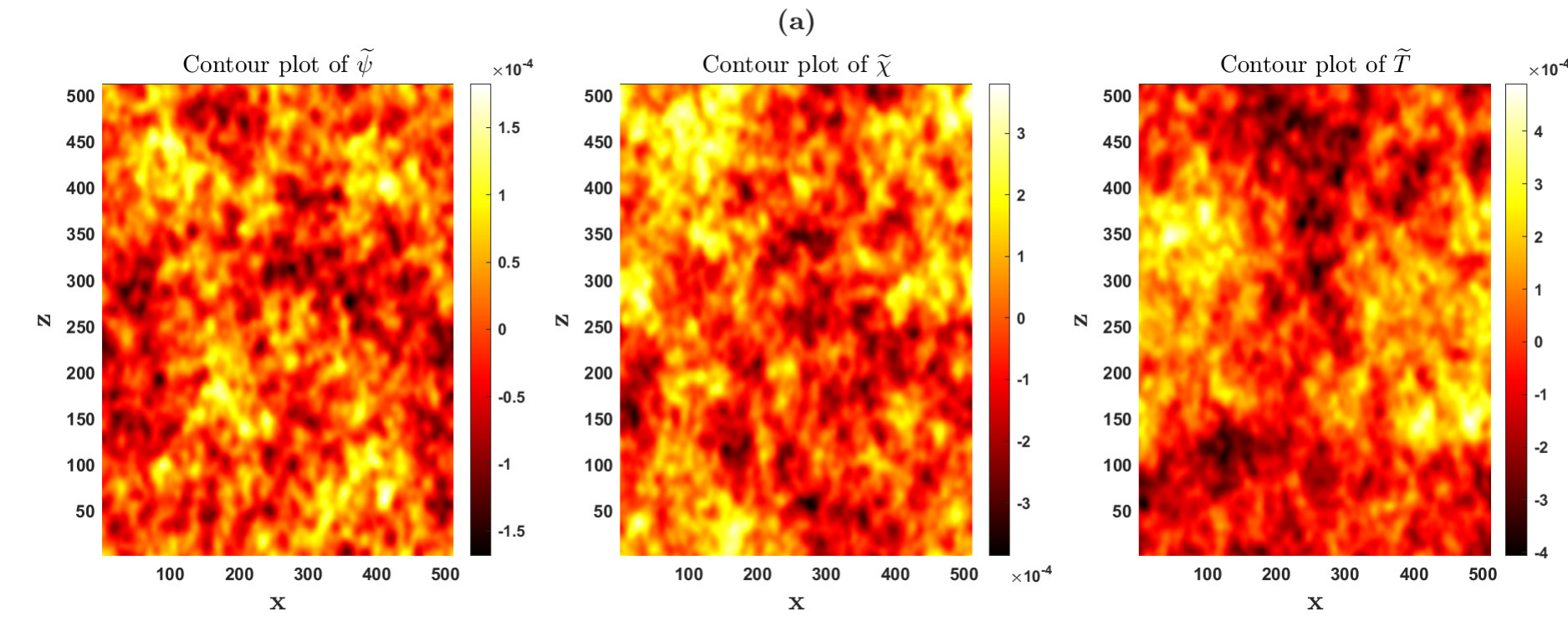}
		\includegraphics[width=6.5in, height=2.12in]{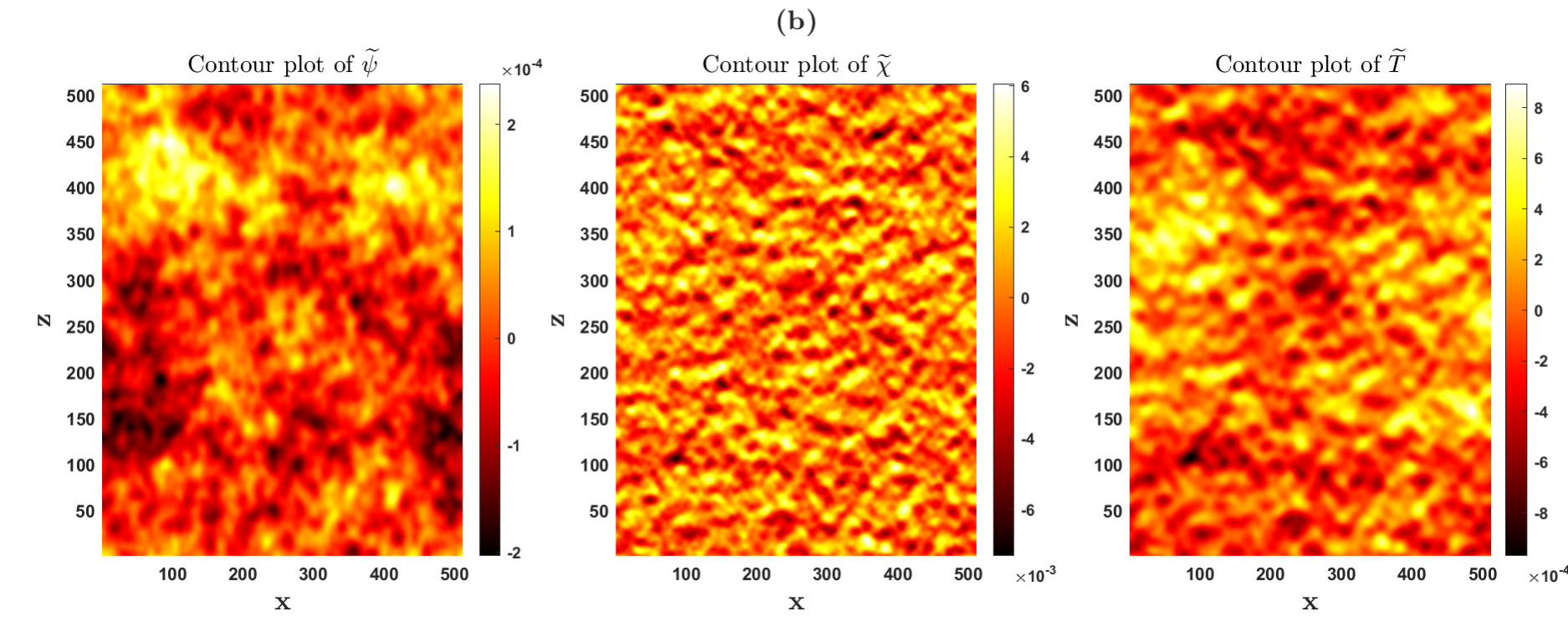}
		\includegraphics[width=6.5in, height=2.12in]{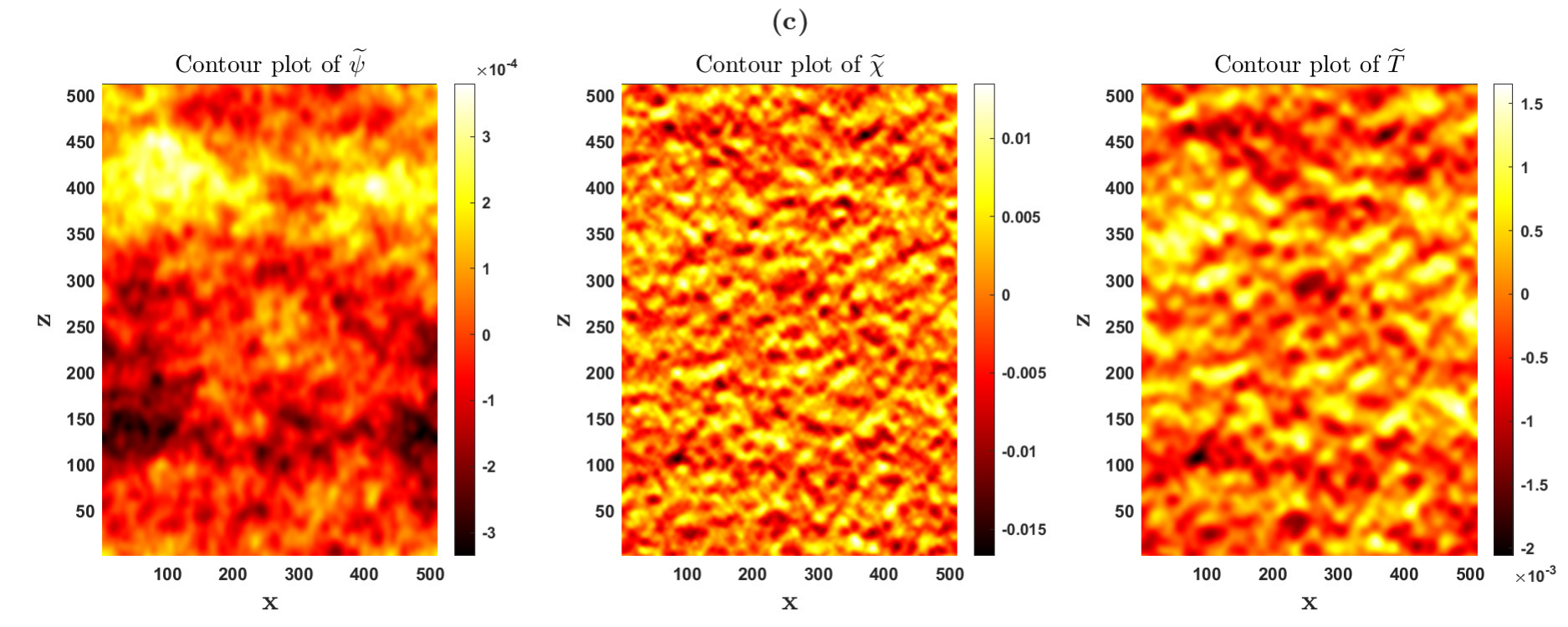}
		\includegraphics[width=6.5in, height=2.12in]{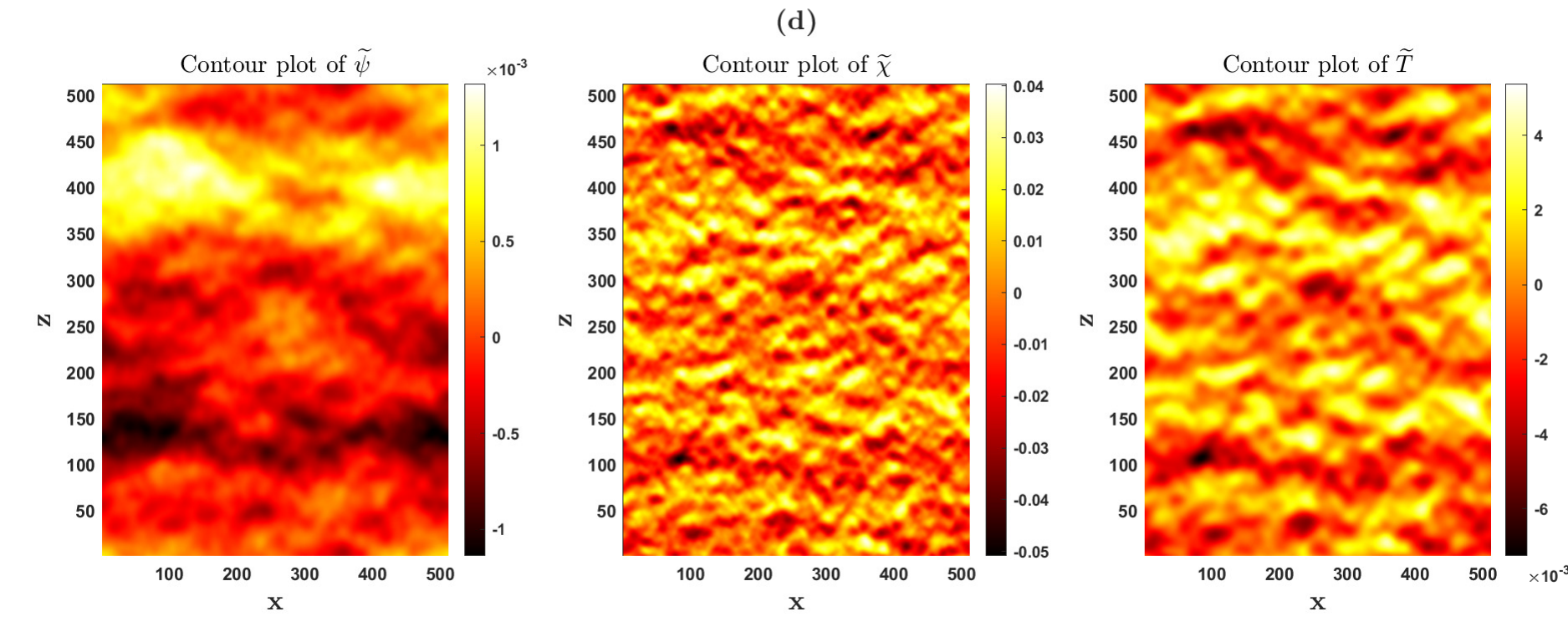}
		\caption{Evolution of the potential function ($\widetilde{\psi}$), density fluctuation ($\widetilde{\chi}$), and temperature ($\widetilde{T}$) at different times (a) $t = 0$, (b) $t = 5$, (c) $t = 10$, and (d) $t = 20$ is shown [Simulation of Eqs. \eqref{eq-mom5}-\eqref{eq-temp5}]. The subplots for $\widetilde{\psi}$ (left panel) illustrate the emergence of large-scale structures in the potential field due to an inverse cascade process. The middle and right panels highlight the development of small-scale eddies, resulting from the forward cascades of density $(\widetilde{\chi})$ and temperature  $(\widetilde{T})$ fluctuations. The parameter values are as in Table \ref{tab-parameter} for the troposphere.}
		\label{fig-tropo}
	\end{figure*}

 
	In the numerical simulation, we consider the parameter values as \cite{Kaladze2023thermal}  $\overline{T} = 255$ K, $\beta = 0.0041$, and $C_0=d\overline{T}/dz<0$ with $(1/\overline{T})(d\overline{T}/dz)=-0.13$. The other parameter values are as in Table \ref{tab-parameter} for the tropospheric region. We show the evolutions of the stream function (associated with the velocity perturbation) and the density and temperature perturbations as contour plots in the $xz$-plane in Fig. \ref{fig-tropo}. Initially, at $t=0$, there is no flow or mean field associated with the perturbations [Subplots (a)], and during the early phase of the simulation, the gravity and thermal modes interact linearly. As the time progresses and the modes gain higher amplitudes, they begin to interact nonlinearly [See subplots (b)-(d) at different times]. During the nonlinear interactions of IGWs and thermal modes, we observe the formation of both large and small-scale structures. Typically, the velocity potential $(\widetilde{\psi})$ tends to cascade large-scale structures (See the left panels) through the process of instability of IGWs. The latter also influences the formation of small-scale structures for the low-frequency density perturbations and the thermal mode of temperature perturbations (See the middle and right panels). These interactions lead to energy transfer among the modes or eddies at different scales (from large to small scales or vice versa), influencing the development of turbulence with time. From the energy spectra, to be shown shortly, we will see the emergence of turbulent states of the fluid flow at different times $t=5$, $10$, $15$, and $20$. We note that the inverse energy transfer (from small to large scales) may be consistent with the fact that IGWs tend to accumulate energy at larger horizontal scales, and that the coexistence of small- and large-scale structures is common in various two-dimensional (2D) turbulence systems, e.g., drift-wave turbulence \citep{ko2003}.
	\par
\begin{figure}[!ht]
		\includegraphics[width=3.5in, height=2.5in]{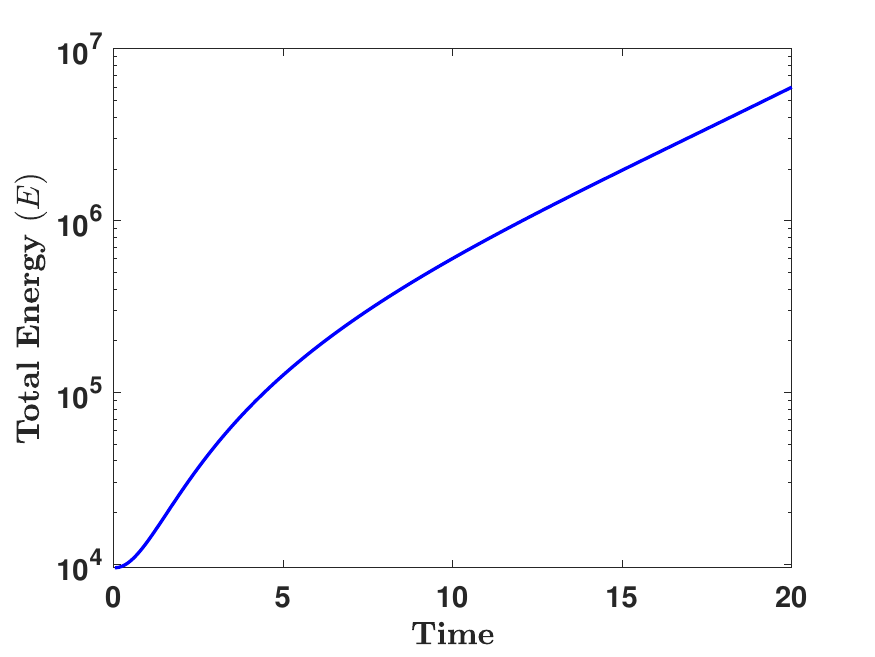}
		\caption{Evolution of the total energy $E$ with time $t$ [Eq. \ref{eq-energy}] is shown. A rapid transfer of energy takes place in the evolution at a later stage of the time period. Such a rapid transfer of energy is due to the instability of IGWs by the effects of the thermal expansion and thermal feedback of the medium to the temperature and density perturbations.  The parameter values are as in Table \ref{tab-parameter} for the troposphere.}
		\label{fig:energy_1}
	\end{figure}

\begin{figure*}[!ht]
		\includegraphics[width=6.5in, height=3in]{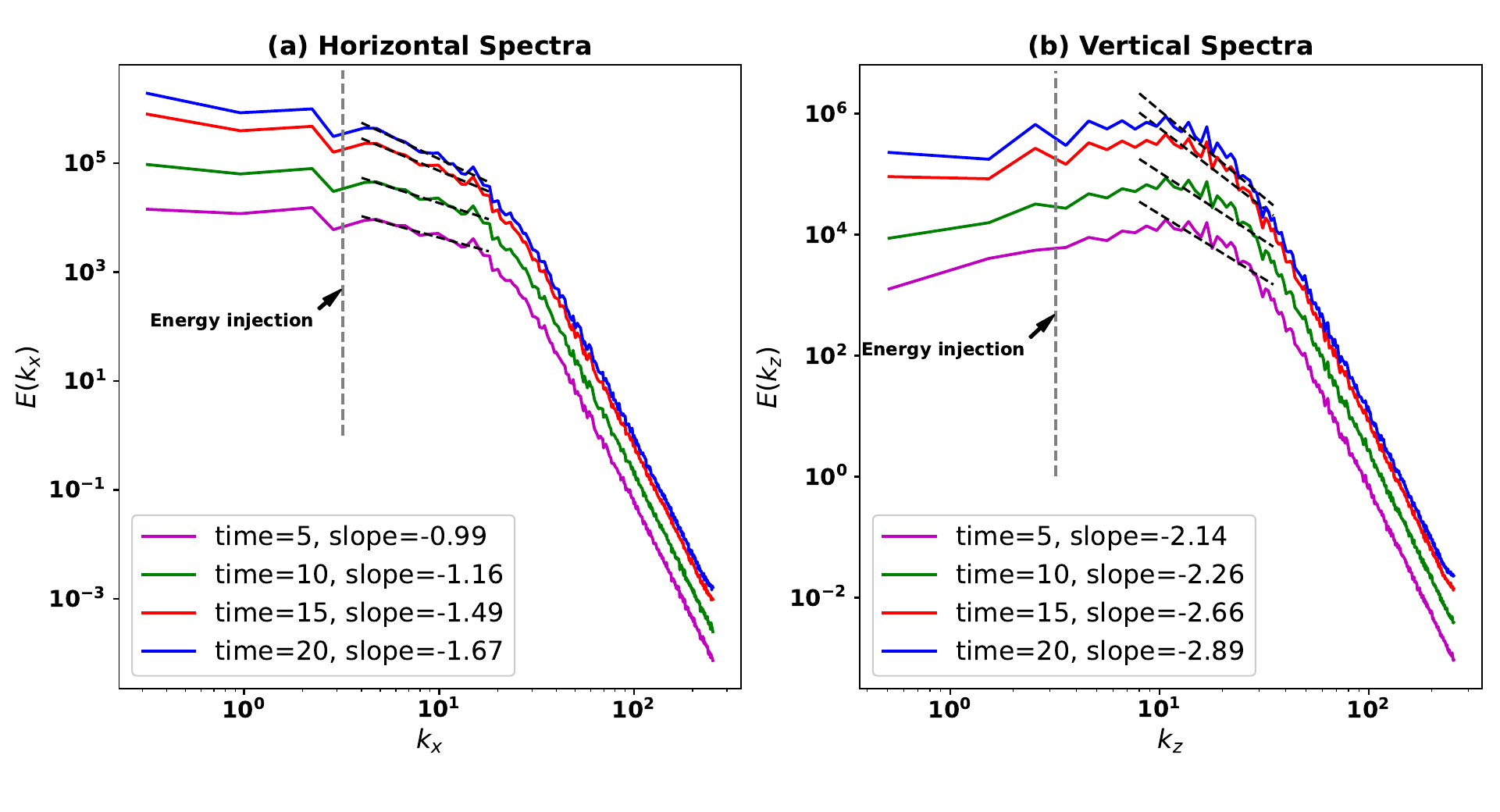}
		\caption{The wave energy spectra $E(k_x)$ and $E(k_z)$ corresponding to the horizontal and vertical wavenumbers are shown at different times $t = 5, 10, 15,$ and $20$. The spectra show a power-law behavior in the inertial range, following $k_x^{-1.67}$ and $k_z^{-2.89}$ for the horizontal and vertical spectra, respectively. The parameter values are the same as given in Table~\ref{tab-parameter} for the troposphere.}
				\label{fig:spectrum_1}
	\end{figure*}

 
	To understand the evolution of turbulent energy, i.e., how fast energy gets redistributed across different scales in the process of energy cascade, where energy transfer occurs from larger to smaller eddies, we plot the total energy [Eq. \eqref{eq-energy}] associated with the evolution of velocity potential and density and temperature fluctuations in the time intervals   $0<t<20$, as shown in Fig. \ref{fig:energy_1}. We observe that in contrast to the IGW turbulence without the influence of temperature perturbations \cite{ShaikhAGW2008}, the energy at the initial stage slowly increases with time. However, as the nonlinear effects intervene, it grows fast with time, but can reach a steady state after a longer period of time. It indicates that as time progresses, the fluid flow becomes more chaotic but less predictable, with larger and more active eddies forming and interacting.  
	As the total energy increases over time, we analyze the energy spectrum at different times $t=5$, $10$, $15$ and $20$.  
	To examine the energy distribution across different sizes (or wave numbers) of eddies, we plot the energy spectra with the horizontal $(k_x)$ and vertical $(k_z)$ wave numbers   as shown in Fig. \ref{fig:spectrum_1}. In the fully developed turbulent state, the horizontal spectra exhibit a slope, $E(k_x)\sim k_x^{-1.67}$ and the vertical spectra follow $E(k_z)\sim k_z^{-2.89}$ (Fig. \ref{fig:spectrum_1}). Here, we note that the right-hand tails of the spectra correspond to the dissipation range, where the power-law scales as $\sim k^{-7}$.
	 These horizontal and vertical spectra are consistent with the theoretical prediction by Kraichnan \cite{kraichnan1967} and the observed wave number spectra for stratified tropospheric (upper region) turbulence without the effects of temperature gradient \cite{nastrom1985}.
	 These distinct slopes also reflect the anisotropic nature of the turbulent flow and the differing energy transfer mechanisms in the horizontal and vertical directions. 
	Thus, in the inertial range of eddies, where the energy cascade is dominant, the energy transfer rate from large to small-scale eddies becomes faster, and the system is said to reach a fully developed turbulent state.
	\par
\begin{figure}[!ht]
		\includegraphics[width=3.5in, height=2.5in]{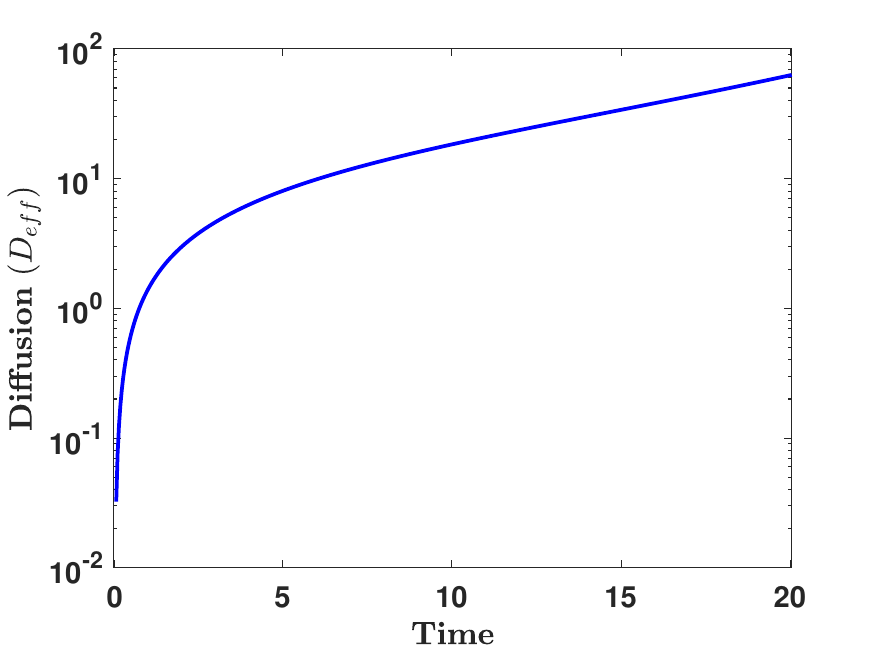}
		\caption{Evolution of the effective diffusion coefficient $D_{\rm{eff}}$ with time $t$ [Eq. \ref{eq-diffusion}] is shown. A rapid increase of the diffusion coefficient over time is a strong indicator of turbulent flow, which promotes rapid transport of momentum, heat, and mass of fluid flow. The parameter values are as in Table \ref{tab-parameter} for the troposphere.}
		\label{fig-diffu}
	\end{figure}

	Gravity-wave turbulence plays a crucial role in vertical mixing, thereby influencing the transport of heat, momentum, or other atmospheric constituents. The effective diffusion coefficient quantifies the enhanced transport rate of the mixing. We estimate the cross-field turbulence transport associated with the self-consistent evolution of small- and large-scale fluctuations. An effective diffusion coefficient depends on momentum transfer, and we calculate it from the following equation. 
	\begin{equation}
		D_{\rm{eff}}=\int_{0}^{\infty}\left\langle\mathbf{v}(\mathbf{r},t)\cdot\mathbf{v}(\mathbf{r},t+t^\prime)\right\rangle dt^\prime,
		\label{eq-diffusion}
	\end{equation} 
	where $\mathbf{v}$ is the fluid velocity. Also, the angular brackets represent the spatial averages, and the ensemble averages are normalized to the unit mass \cite{ShaikhAGW2008}. In our simulation,  we measure the effective diffusion coefficient $D_{\rm{eff}}$ for the velocity potential field, associated with the large-scale flows. Initially, the field perturbation is Gaussian, in which case, the transport is lower. In the later stages, the enhanced cross-field diffusion coefficient in our simulation proves consistent with the development of turbulent energy (See Fig. \ref{fig:energy_1}) and large-scale flows (See Fig. \ref{fig-diffu}) \cite{ShaikhAGW2008}. Such a high diffusion coefficient is an indication of enhanced vertical mixing due to instability of IGWs coupled to thermal modes, leading to an increased energy transport between different layers in the troposphere. However, in the steady state, the nonlinearly coupled IGW and thermal modes can tend to form stationary structures and the diffusion coefficient $D_{\rm{eff}}$ can saturate eventually after a longer period of time. 
	\subsection{Turbulence spectra in the stratosphere} \label{sec-strato}
\begin{figure*}[!ht]
		\includegraphics[width=6.5in, height=2in]{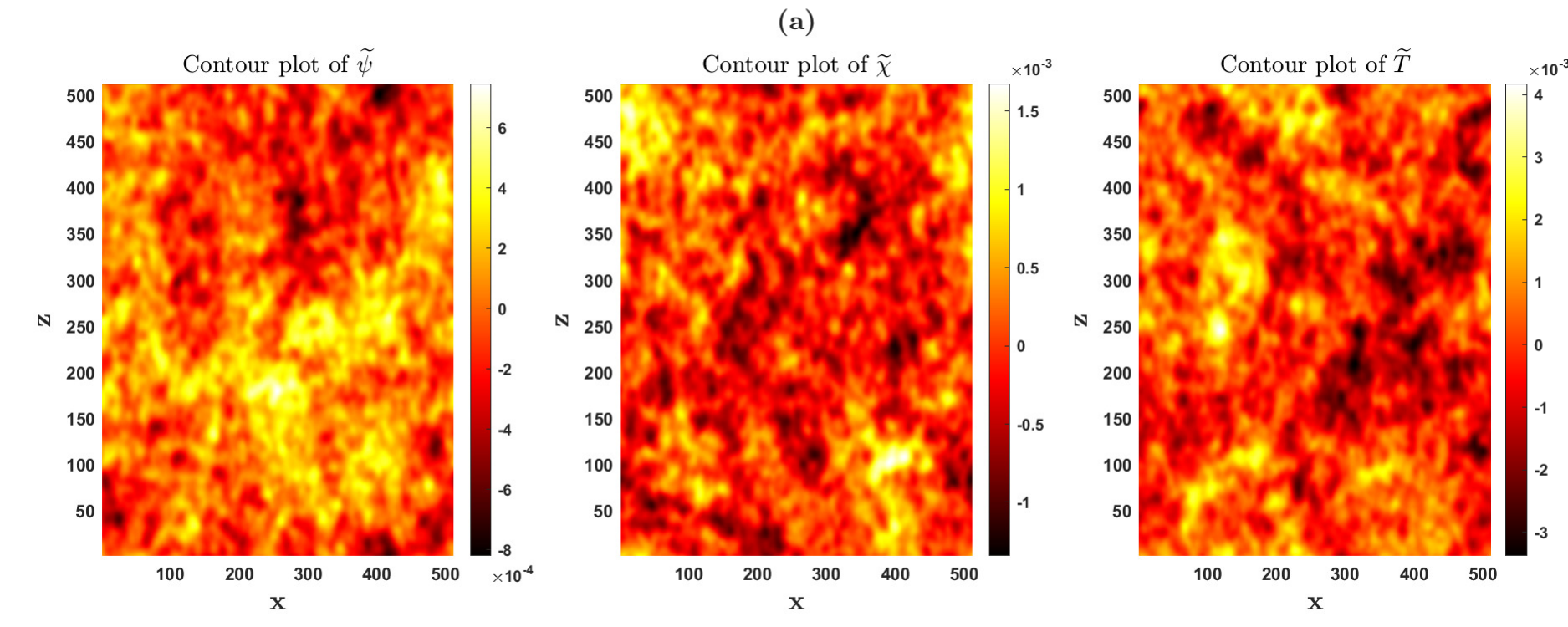}
		\includegraphics[width=6.5in, height=2in]{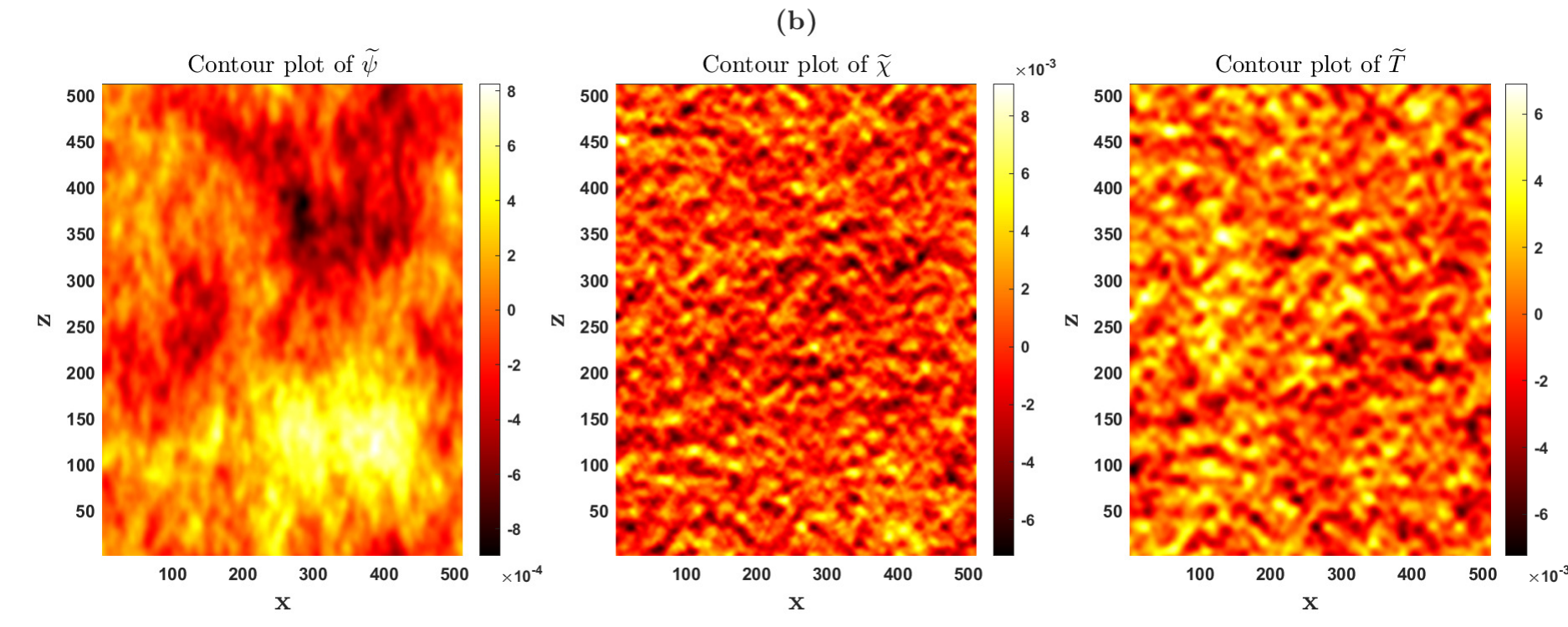}
		\includegraphics[width=6.5in, height=2in]{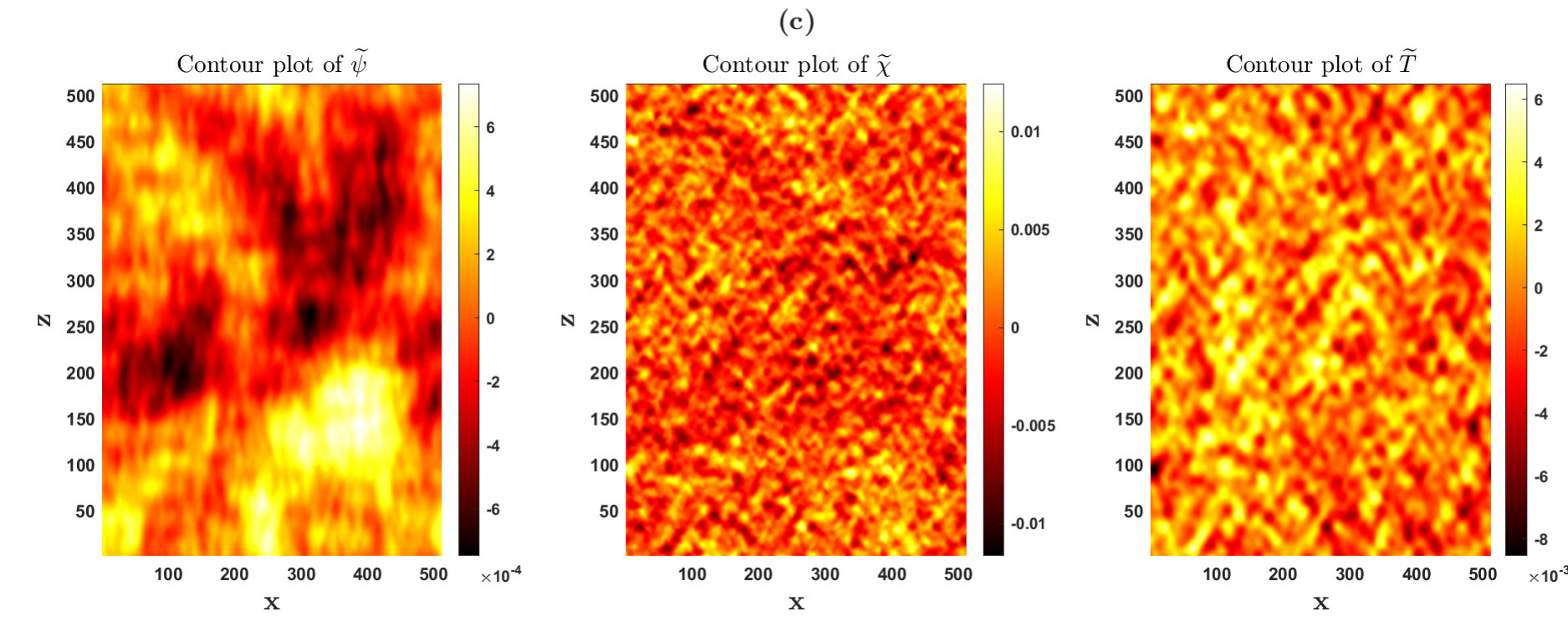}
		\includegraphics[width=6.5in, height=2in]{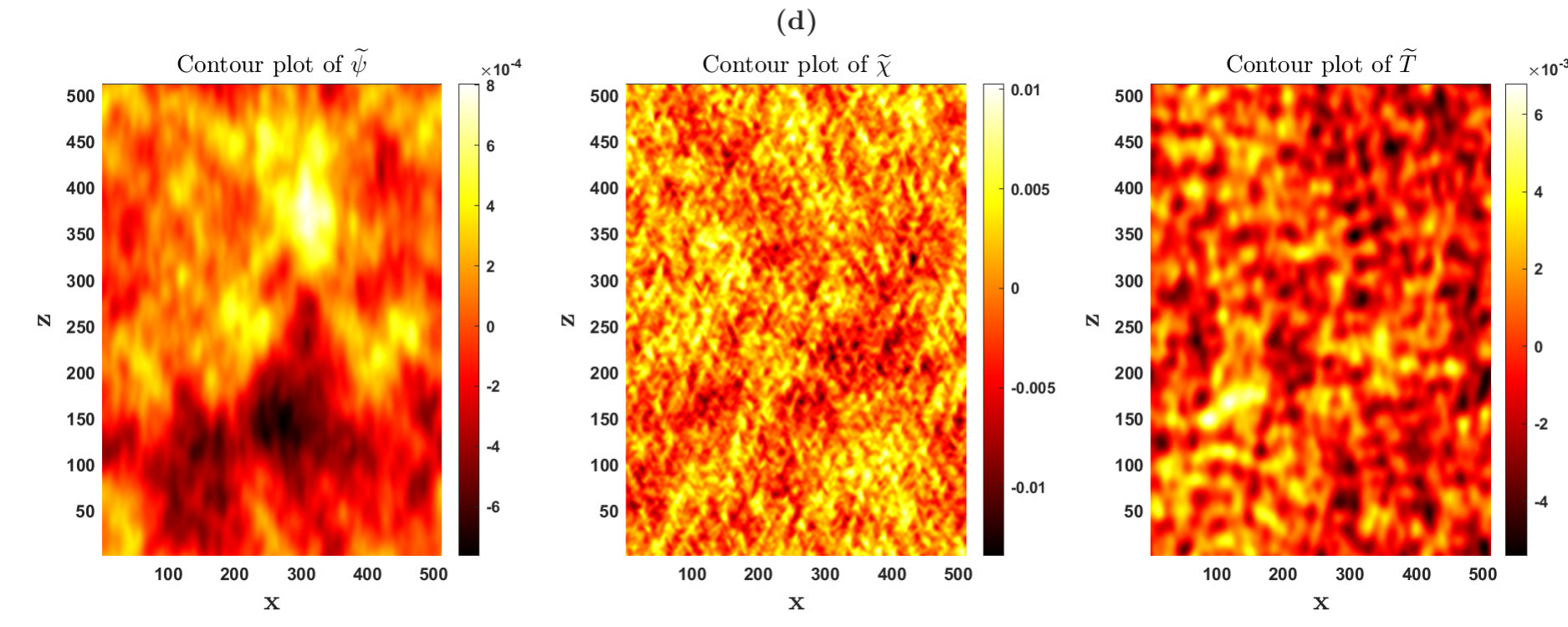}
		\caption{Evolution of the potential function ($\widetilde{\psi}$, left panel), density fluctuation  ($\widetilde{\chi}$, middle panel), and temperature ($\widetilde{T}$, right panel) at different times, $t = 0$ [plot (a)], $t = 5$ [plot (b)], $t = 10$ [plot (c)], and $t = 20$ [plot(d)] is shown. Simulation results of Eqs. \eqref{eq-mom5}--\eqref{eq-temp5} from a random initial state show the formation of large-scale structures in the potential fluctuations due to the inverse cascade process, while small-scale structures form in the density fluctuations and temperature variations due to the forward cascade processes. The parameter values are as in Table \ref{tab-parameter} for the stratosphere.}
		\label{fig-strato}
	\end{figure*}

	We turn to investigate the turbulent spectra in the stratospheric region where $C_0\equiv d\overline{T}/dz > 0$. In this region, the stratified flows are linearly stable in the absence of thermal feedback \cite{Kaladze2024temperature}. While investigating the vortex motion of IGWs, we have observed that the linear IGW mode can become unstable due to coupling with the thermal wave, and the influence of the thermal feedback of the stratified fluid on density and temperature fluctuations. However, the instability saturates, and the corresponding growth rate can vanish due to the strong influence of the thermal diffusion [\textit{cf}. Fig. \ref{fig2-disp} (b)]. Thus, although the instability of IGWs can occur, in contrast to the tropospheric flows, full development of turbulence may not occur in the stratosphere. This section aims to explore the turbulent spectra and verify the theory \cite{kraichnan1967}, similar to Sec. \ref{sec-tropo} but limited to relevant details. For the stratospheric region ($15 < z < 50$ km), we consider the temperature as $\overline{T} = 226$ K, with a thermal expansion coefficient $\beta = 0.0045$, and a positive temperature gradient ($dT_0/dz > 0$), resulting $(1/\overline{T})(d\overline{T}/dz)=0.37$. The other parameters are as in Table \ref{tab-parameter}.  We present the simulation results for the nonlinear evolution of the velocity potential, density, and temperature fluctuations at different time intervals as in Fig. \ref{fig-strato}.  In this case, we also observe large-scale flows of the velocity potential field (left panel)and small-scale structures for density (middle panel) and temperature (right panel) fluctuations.
	\par
\begin{figure}[!ht]
		\includegraphics[width=3.5in, height=2.5in]{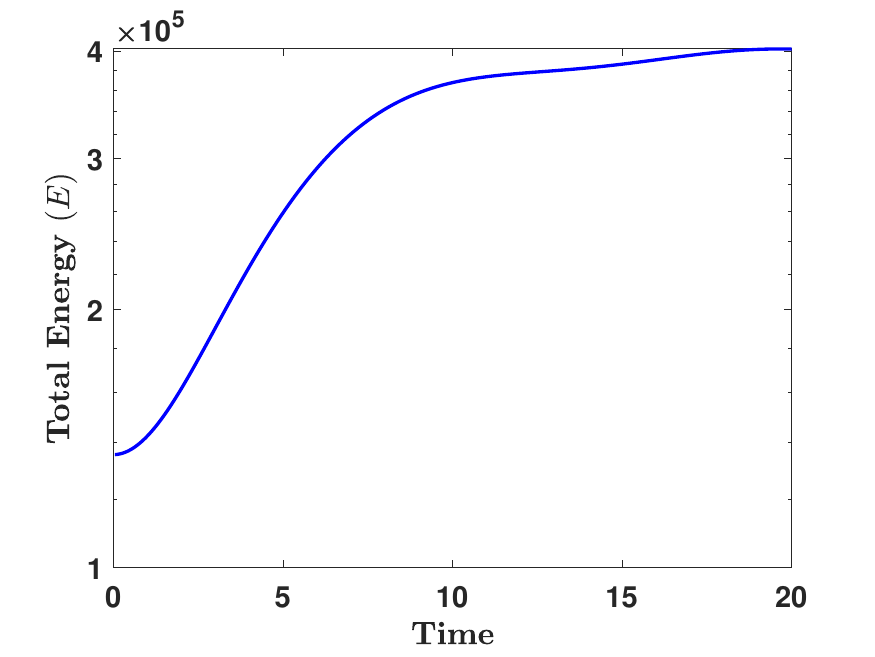}
		\caption{The total energy $E$ [Eq. \eqref{eq-energy}] in the stratospheric region, showing saturation at later times. The parameter values are as in Table \ref{tab-parameter} for the stratosphere.}
		\label{fig-energy2}
	\end{figure}
\begin{figure*}[!ht]
		\includegraphics[width=6.5in, height=3in]{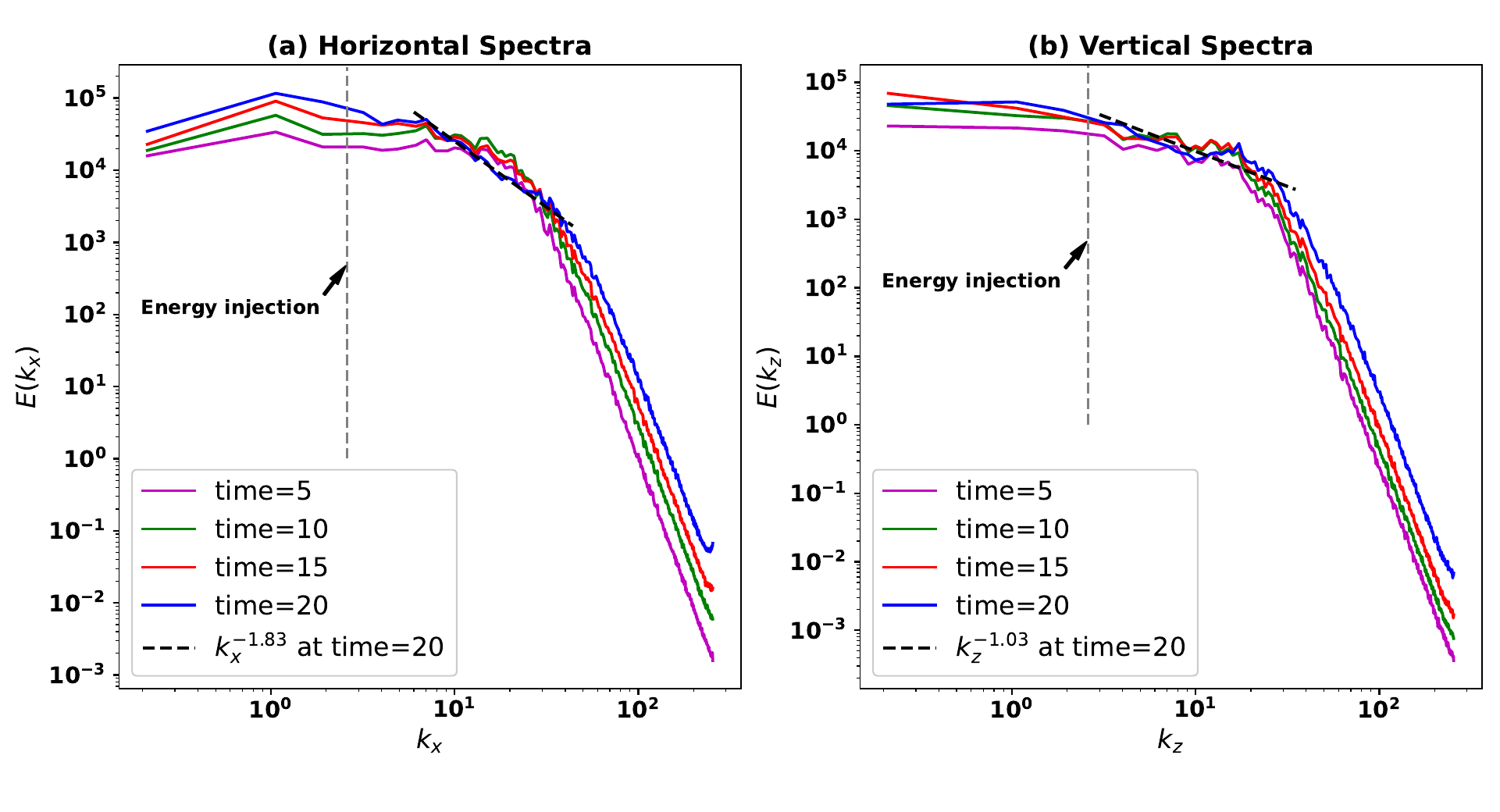}
		\caption{The horiozontal and vertical energy spectra $E(k_x)$ and $E(k_z)$, showing power-law scaling close to $k_x^{-1.83}$ and $k_z^{-1.03}$, respectively in the inertial range for the stratospheric region. The parameter values are as in Table~\ref{tab-parameter} for the stratosphere.}
		\label{fig-spectrum3}
	\end{figure*}

	Figure \ref{fig-energy2} shows the evolution of total energy over time. In contrast to the tropospheric region, although energy increases over an initial period, it eventually reaches a steady state within the same time interval without significant growth due to saturation of nonlinear interactions. It follows that the IGW amplitude can no longer grow to trigger strong instabilities; as a result, the energy transfer from large to small scales through turbulence saturates. Such a saturation is consistent with the linear analysis in Sec. \ref{sec-basic}, where we have seen that for a fixed vertical size, the instability growth rate does not increase with an increase of horizontal wave number $k_x$ unless the thermal feedback of the medium on the density and temperature perturbations is sufficiently strong [\textit{cf}. Fig. \ref{fig2-disp}(b)]. Thus, in stratospheric stratified flows with positive temperature gradient, the nonlinear saturation limits the further growth of thermoacoustic IGW energy at larger scales, i.e., turbulence may not fully develop. We will clarify it from the analysis of energy spectra.   
	As shown in Fig. \ref{fig-spectrum3}, the power spectra associated with the turbulence exhibit a power law $E(k_x) \sim k_x^{-1.83}$ for horizontal flows and $E(k_z) \sim k_z^{-1.03}$ for vertical flows, indicating that turbulent eddies are the dominant processes governing spectral transfer. 
Here, we note that the right-hand tail of the spectra corresponds to the dissipation range, where the power-law scales as $\sim k^{-7}$. However, these horizontal and vertical spectra differ from the theoretical prediction by Kraichnan \cite{kraichnan1967} and the observed wave number spectra for stratified stratospheric (lower region) turbulence without the effects of temperature gradient \cite{nastrom1985}.	 
	\par
\begin{figure}[!ht]
		\includegraphics[width=3.5in, height=2.5in]{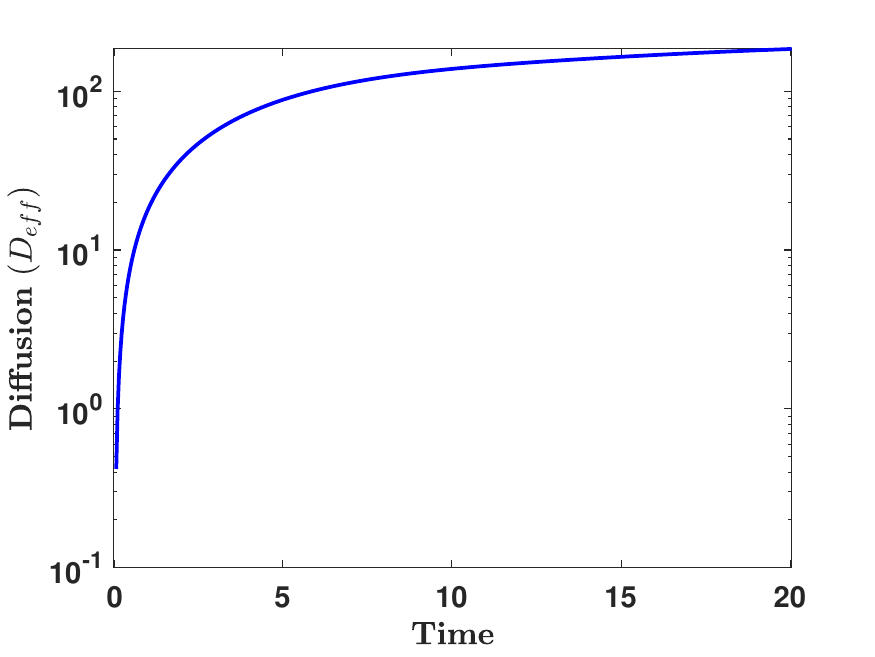}
		\caption{Time evolution of the effective diffusion coefficient $D_{\rm{eff}}$ [Eq. \eqref{eq-diffusion}], representing the diffusion driven by large-scale velocity potential and small-scale variations in density and temperature. Initially, the transport of heat and momentum is higher, but it eventually reaches a steady state due to nonlinear saturation. The parameter values are as in Table \ref{tab-parameter} for the stratosphere.}
		\label{fig-diffusion2}
	\end{figure}

 
	Finally, we estimate the turbulent transport coefficient (See Fig. \ref{fig-diffusion2}) in the same way as in the case of tropospheric turbulence (Sec. \ref{sec-tropo}) using Eq. \eqref{eq-diffusion}. The effective diffusion coefficient ($D_{\text{eff}}$) shows a rapid increase during the early stages of the simulation. However, after a period, $D_{\text{eff}}$ begins to reach a steady state due to nonlinear saturation, i.e., when IGW amplitude reaches a limit and the instability growth rate saturates or tends to vanish at a finite wave number. In this situation, energy transfer from large to small scales gets significantly reduced, which is consistent with the evolution of the total energy (Fig. \ref{fig-energy2}).   In this case, the nonlinearly coupled IGW and thermal modes also tend to form stationary structures. However, the diffusion coefficient $D_{\rm{eff}}$ saturates after a shorter period of time compared to the troposphere (\textit{cf}. Fig. \ref{fig-diffu}).   
	\begin{table}[!ht]
		\centering
		\begin{tabular}{|c|c|c|}
			\hline
			Parameters&Troposphere & Stratosphere\\
			\hline
			$\alpha^\prime$ & $0.98$  & $0.91$ \\
			\hline
			$\zeta^\prime$ & $0.97$ & $0.97$ \\
			\hline
			Thermal expansion ($\beta$)&$0.0041~(\rm{K}^{-1})$&$0.0045~(\rm{K}^{-1})$\\
			\hline
			Temperature $\left(\overline{T}\right)$ &$255$ ($K$)&$226$ ($K$)\\
			\hline
			$\kappa^\prime$&$0.00085$&$0.000279$\\
			\hline
			$A_1$&$0.014$&$0.0356$\\
			\hline
			$A_2$&$0.015$&$0.012$\\
			\hline
			$ \frac{1}{\overline{T}}\frac{d\overline{T}}{dz}$&$-0.13$&$0.37$\\
			\hline
		\end{tabular}
		\caption{Sets of parameter values considered in the simulations for the tropospheric and stratospheric regions are shown. We have taken data from the open source ``U.S. Standard Atmosphere Air Properties" \cite{USdata}. Some more details are in Ref. \cite{Kaladze2023thermal}. We take the time and length scales as $\omega^{-1}_g\sim20$ s, $H\sim4\times 10^3$ m. }
		\label{tab-parameter}
	\end{table}
	\section{Conclusion}\label{sec-conclusion}
	We have proposed a two-dimensional atmospheric fluid model for the coupling of internal gravity waves (IGWs) and thermal modes by the influences of the temperature dependent density inhomogeneity, thermal gradient, and thermal feedback on temperature and density fluctuations. We have shown that the equilibrium temperature profile may not be linear, in general, in the presence of a heat source. However, if the heat source is given by the Fourier law, in the linear approximation, the thermal dependence agrees with the linear relation without any heat source \citep{Kaladze2023thermal}. For the two-dimensional (2D) evolution of thermoacoustic solitary vortices, we have derived a set of three coupled nonlinear partial differential equations for the velocity potential field (stream function), and density and temperature fluctuations. In the small-amplitude limit (linear regime), we show that due to coupling with the thermal mode, IGWs can become unstable by the influences of the thermal feedback. While the instability growth rate increases in the tropospheric region  ($0 < z < 15$ km) with negative temperature gradient, the same can have a cut-off in the stratospheric  ($15 < z < 50$ km) stratified fluids with positive temperature gradient unless the thermal feedback is sufficiently high. 
	\par 
	In the nonlinear regime, we perform a simulation approach to study the evolution and characteristics of turbulent flow. We observe that in both the tropospheric and stratospheric fluids, a random initial state evolves into large-scale velocity potential structures due to an inverse cascade, and small-scale features in density and temperature fluctuations emerge from a forward cascade. This inverse cascade behavior is consistent with the dual-cascade dynamics typical of two-dimensional turbulence \cite{ShaikhAGW2008} and may not directly correspond to energy transfer processes in a fully three-dimensional flow. A rapid increase of the wave energy and effective diffusion, 
		and the energy spectra with horizontal slope $k_x^{-1.67}$ and vertical slope $k_z^{-2.89}$, 
	indicate the possibility of fully developed turbulence in the tropospheric stratified flows. 
	These horizontal and vertical wave number spectra are consistent with the theory by Kraichnan \cite{kraichnan1967} and the observed phenomena \cite{nastrom1985} in the troposphere without any temperature gradient effects.  
	On the other hand, stratospheric fluid flows correspond to an energy spectrum 
	with a horizontal slope of $E(k_x)\propto k_x^{-1.83}$ and a vertical slope of $E(k_z)\propto k_z^{-1.03}$, together with steady states of the total wave energy and the diffusion coefficient, indicating a reduced transfer of energy from large to small scales. These horizontal and vertical spectra differ from the theory \cite{kraichnan1967} and the observations \cite{nastrom1985} in the stratosphere without temperature gradient effects.    
	\par
	To conclude, IGWs that couple to thermal modes can propagate as thermoacoustic IGWs in the atmosphere, which have distinct characteristics compared to classical IGWs in the literature. However, as they propagate upwards, their amplitudes can grow due to instability, caused by the thermal feedback of the stratified medium on density and temperature fluctuations. At some point, they break and eventually lead to turbulence due to faster energy transfer from large- to small-scale perturbations. Thus, the evolution of thermoacoustic IGWs and the vortex turbulent motion of stratified fluids can significantly affect momentum and heat transfer \cite{Rogachevskii2024}, as well as influence the formation of severe weather patterns and atmospheric conditions.  
	\newpage
	\bibliographystyle{apsrev4-2} 
	\bibliography{Reference}

\begin{thebibliography}{28}%
\makeatletter
\providecommand \@ifxundefined [1]{%
 \@ifx{#1\undefined}
}%
\providecommand \@ifnum [1]{%
 \ifnum #1\expandafter \@firstoftwo
 \else \expandafter \@secondoftwo
 \fi
}%
\providecommand \@ifx [1]{%
 \ifx #1\expandafter \@firstoftwo
 \else \expandafter \@secondoftwo
 \fi
}%
\providecommand \natexlab [1]{#1}%
\providecommand \enquote  [1]{``#1''}%
\providecommand \bibnamefont  [1]{#1}%
\providecommand \bibfnamefont [1]{#1}%
\providecommand \citenamefont [1]{#1}%
\providecommand \href@noop [0]{\@secondoftwo}%
\providecommand \href [0]{\begingroup \@sanitize@url \@href}%
\providecommand \@href[1]{\@@startlink{#1}\@@href}%
\providecommand \@@href[1]{\endgroup#1\@@endlink}%
\providecommand \@sanitize@url [0]{\catcode `\\12\catcode `\$12\catcode
  `\&12\catcode `\#12\catcode `\^12\catcode `\_12\catcode `\%12\relax}%
\providecommand \@@startlink[1]{}%
\providecommand \@@endlink[0]{}%
\providecommand \url  [0]{\begingroup\@sanitize@url \@url }%
\providecommand \@url [1]{\endgroup\@href {#1}{\urlprefix }}%
\providecommand \urlprefix  [0]{URL }%
\providecommand \Eprint [0]{\href }%
\providecommand \doibase [0]{https://doi.org/}%
\providecommand \selectlanguage [0]{\@gobble}%
\providecommand \bibinfo  [0]{\@secondoftwo}%
\providecommand \bibfield  [0]{\@secondoftwo}%
\providecommand \translation [1]{[#1]}%
\providecommand \BibitemOpen [0]{}%
\providecommand \bibitemStop [0]{}%
\providecommand \bibitemNoStop [0]{.\EOS\space}%
\providecommand \EOS [0]{\spacefactor3000\relax}%
\providecommand \BibitemShut  [1]{\csname bibitem#1\endcsname}%
\let\auto@bib@innerbib\@empty
\bibitem [{\citenamefont {Busse}(1977)}]{Busse1977}%
  \BibitemOpen
  \bibfield  {author} {\bibinfo {author} {\bibfnamefont {F.~H.}\ \bibnamefont
  {Busse}},\ }\href {https://doi.org/10.1017/S0022112077210986} {\bibfield
  {journal} {\bibinfo  {journal} {Journal of Fluid Mechanics}\ }\textbf
  {\bibinfo {volume} {82}},\ \bibinfo {pages} {794} (\bibinfo {year}
  {1977})}\BibitemShut {NoStop}%
\bibitem [{\citenamefont {Acheson}\ and\ \citenamefont
  {Hide}(1973)}]{DJAcheson_1973}%
  \BibitemOpen
  \bibfield  {author} {\bibinfo {author} {\bibfnamefont {D.~J.}\ \bibnamefont
  {Acheson}}\ and\ \bibinfo {author} {\bibfnamefont {R.}~\bibnamefont {Hide}},\
  }\href {https://doi.org/10.1088/0034-4885/36/2/002} {\bibfield  {journal}
  {\bibinfo  {journal} {Reports on Progress in Physics}\ }\textbf {\bibinfo
  {volume} {36}},\ \bibinfo {pages} {159} (\bibinfo {year} {1973})}\BibitemShut
  {NoStop}%
\bibitem [{\citenamefont {Miyoshi}\ and\ \citenamefont
  {Fujiwara}(2008)}]{Miyoshi2008Gravity}%
  \BibitemOpen
  \bibfield  {author} {\bibinfo {author} {\bibfnamefont {Y.}~\bibnamefont
  {Miyoshi}}\ and\ \bibinfo {author} {\bibfnamefont {H.}~\bibnamefont
  {Fujiwara}},\ }\href {https://doi.org/https://doi.org/10.1029/2007JD008874}
  {\bibfield  {journal} {\bibinfo  {journal} {Journal of Geophysical Research:
  Atmospheres}\ }\textbf {\bibinfo {volume} {113}},\ \bibinfo {pages} {D01101}
  (\bibinfo {year} {2008})}\BibitemShut {NoStop}%
\bibitem [{\citenamefont {Plougonven}\ and\ \citenamefont
  {Zhang}(2014)}]{Plougonven2014IGW}%
  \BibitemOpen
  \bibfield  {author} {\bibinfo {author} {\bibfnamefont {R.}~\bibnamefont
  {Plougonven}}\ and\ \bibinfo {author} {\bibfnamefont {F.}~\bibnamefont
  {Zhang}},\ }\href {https://doi.org/https://doi.org/10.1002/2012RG000419}
  {\bibfield  {journal} {\bibinfo  {journal} {Reviews of Geophysics}\ }\textbf
  {\bibinfo {volume} {52}},\ \bibinfo {pages} {33} (\bibinfo {year}
  {2014})}\BibitemShut {NoStop}%
\bibitem [{\citenamefont {Kaladze}\ \emph {et~al.}(2008)\citenamefont
  {Kaladze}, \citenamefont {Pokhotelov}, \citenamefont {Shah}, \citenamefont
  {Khan},\ and\ \citenamefont {Stenflo}}]{Kaladze2008acoustics}%
  \BibitemOpen
  \bibfield  {author} {\bibinfo {author} {\bibfnamefont {T.~D.}\ \bibnamefont
  {Kaladze}}, \bibinfo {author} {\bibfnamefont {O.~A.}\ \bibnamefont
  {Pokhotelov}}, \bibinfo {author} {\bibfnamefont {H.~A.}\ \bibnamefont
  {Shah}}, \bibinfo {author} {\bibfnamefont {M.~I.}\ \bibnamefont {Khan}},\
  and\ \bibinfo {author} {\bibfnamefont {L.}~\bibnamefont {Stenflo}},\ }\href
  {https://doi.org/https://doi.org/10.1016/j.jastp.2008.06.009} {\bibfield
  {journal} {\bibinfo  {journal} {Journal of Atmospheric and Solar-Terrestrial
  Physics}\ }\textbf {\bibinfo {volume} {70}},\ \bibinfo {pages} {1607}
  (\bibinfo {year} {2008})}\BibitemShut {NoStop}%
\bibitem [{\citenamefont {Chen}(1983)}]{chen1983}%
  \BibitemOpen
  \bibfield  {author} {\bibinfo {author} {\bibfnamefont {F.~F.}\ \bibnamefont
  {Chen}},\ }\href@noop {} {\emph {\bibinfo {title} {Introduction to Plasma
  Physics and Controlled Fusion}}},\ Vol.~\bibinfo {volume} {1}\ (\bibinfo
  {publisher} {Plenum Press, New York and London, Second Edition},\ \bibinfo
  {year} {1983})\ pp.\ \bibinfo {pages} {95--114}\BibitemShut {NoStop}%
\bibitem [{\citenamefont {Shaikh}\ \emph {et~al.}(2008)\citenamefont {Shaikh},
  \citenamefont {Shukla},\ and\ \citenamefont {Stenflo}}]{ShaikhAGW2008}%
  \BibitemOpen
  \bibfield  {author} {\bibinfo {author} {\bibfnamefont {D.}~\bibnamefont
  {Shaikh}}, \bibinfo {author} {\bibfnamefont {P.~K.}\ \bibnamefont {Shukla}},\
  and\ \bibinfo {author} {\bibfnamefont {L.}~\bibnamefont {Stenflo}},\ }\href
  {https://doi.org/https://doi.org/10.1029/2007JD009305} {\bibfield  {journal}
  {\bibinfo  {journal} {Journal of Geophysical Research: Atmospheres}\ }\textbf
  {\bibinfo {volume} {113}},\ \bibinfo {pages} {D06108} (\bibinfo {year}
  {2008})}\BibitemShut {NoStop}%
\bibitem [{\citenamefont {Stenflo}(1996)}]{Stenflo1996}%
  \BibitemOpen
  \bibfield  {author} {\bibinfo {author} {\bibfnamefont {L.}~\bibnamefont
  {Stenflo}},\ }\href {https://doi.org/10.1016/S0375-9601(96)00671-8}
  {\bibfield  {journal} {\bibinfo  {journal} {Physics Letters A}\ }\textbf
  {\bibinfo {volume} {222}},\ \bibinfo {pages} {378} (\bibinfo {year}
  {1996})}\BibitemShut {NoStop}%
\bibitem [{\citenamefont {Kaladze}\ and\ \citenamefont
  {Misra}(2024)}]{Kaladze2024temperature}%
  \BibitemOpen
  \bibfield  {author} {\bibinfo {author} {\bibfnamefont {T.~D.}\ \bibnamefont
  {Kaladze}}\ and\ \bibinfo {author} {\bibfnamefont {A.~P.}\ \bibnamefont
  {Misra}},\ }\href {https://doi.org/10.1088/1402-4896/ad5ccc} {\bibfield
  {journal} {\bibinfo  {journal} {Physica Scripta}\ }\textbf {\bibinfo {volume}
  {99}},\ \bibinfo {pages} {085013} (\bibinfo {year} {2024})}\BibitemShut
  {NoStop}%
\bibitem [{\citenamefont {Misra}\ and\ \citenamefont
  {Banerjee}(2025)}]{misra2025}%
  \BibitemOpen
  \bibfield  {author} {\bibinfo {author} {\bibfnamefont {A.~P.}\ \bibnamefont
  {Misra}}\ and\ \bibinfo {author} {\bibfnamefont {G.}~\bibnamefont
  {Banerjee}},\ }\href {https://doi.org/10.1016/j.wavemoti.2024.103451}
  {\bibfield  {journal} {\bibinfo  {journal} {Wave Motion}\ }\textbf {\bibinfo
  {volume} {133}},\ \bibinfo {pages} {103451} (\bibinfo {year}
  {2025})}\BibitemShut {NoStop}%
\bibitem [{\citenamefont {Kryuchkov}\ \emph {et~al.}(2018)\citenamefont
  {Kryuchkov}, \citenamefont {Yakovlev}, \citenamefont {Gorbunov},
  \citenamefont {Cou\"edel}, \citenamefont {Lipaev},\ and\ \citenamefont
  {Yurchenko}}]{kryuchkov2018}%
  \BibitemOpen
  \bibfield  {author} {\bibinfo {author} {\bibfnamefont {N.~P.}\ \bibnamefont
  {Kryuchkov}}, \bibinfo {author} {\bibfnamefont {E.~V.}\ \bibnamefont
  {Yakovlev}}, \bibinfo {author} {\bibfnamefont {E.~A.}\ \bibnamefont
  {Gorbunov}}, \bibinfo {author} {\bibfnamefont {L.}~\bibnamefont {Cou\"edel}},
  \bibinfo {author} {\bibfnamefont {A.~M.}\ \bibnamefont {Lipaev}},\ and\
  \bibinfo {author} {\bibfnamefont {S.~O.}\ \bibnamefont {Yurchenko}},\ }\href
  {https://doi.org/10.1103/PhysRevLett.121.075003} {\bibfield  {journal}
  {\bibinfo  {journal} {Phys. Rev. Lett.}\ }\textbf {\bibinfo {volume} {121}},\
  \bibinfo {pages} {075003} (\bibinfo {year} {2018})}\BibitemShut {NoStop}%
\bibitem [{\citenamefont {Tur}\ and\ \citenamefont
  {Yanovsky}(2013)}]{tur2013non}%
  \BibitemOpen
  \bibfield  {author} {\bibinfo {author} {\bibfnamefont {A.}~\bibnamefont
  {Tur}}\ and\ \bibinfo {author} {\bibfnamefont {V.}~\bibnamefont {Yanovsky}},\
  }\href {https://doi.org/10.4236/ojfd.2013.32009} {\bibfield  {journal}
  {\bibinfo  {journal} {Open Journal of Fluid Dynamics}\ }\textbf {\bibinfo
  {volume} {3}},\ \bibinfo {pages} {64} (\bibinfo {year} {2013})}\BibitemShut
  {NoStop}%
\bibitem [{\citenamefont {Tur}\ \emph {et~al.}(2013)\citenamefont {Tur},
  \citenamefont {Chabane},\ and\ \citenamefont {Yanovsky}}]{tur2013Rorate}%
  \BibitemOpen
  \bibfield  {author} {\bibinfo {author} {\bibfnamefont {A.}~\bibnamefont
  {Tur}}, \bibinfo {author} {\bibfnamefont {M.}~\bibnamefont {Chabane}},\ and\
  \bibinfo {author} {\bibfnamefont {V.}~\bibnamefont {Yanovsky}},\ }\href
  {https://doi.org/10.4236/ojfd.2013.34041} {\bibfield  {journal} {\bibinfo
  {journal} {Open Journal of Fluid Dynamics}\ }\textbf {\bibinfo {volume}
  {03}},\ \bibinfo {pages} {340} (\bibinfo {year} {2013})}\BibitemShut
  {NoStop}%
\bibitem [{\citenamefont {Kopp}\ \emph {et~al.}(2021)\citenamefont {Kopp},
  \citenamefont {Tur},\ and\ \citenamefont {Yanovsky}}]{Kopp2021}%
  \BibitemOpen
  \bibfield  {author} {\bibinfo {author} {\bibfnamefont {M.~I.}\ \bibnamefont
  {Kopp}}, \bibinfo {author} {\bibfnamefont {A.~V.}\ \bibnamefont {Tur}},\ and\
  \bibinfo {author} {\bibfnamefont {V.~V.}\ \bibnamefont {Yanovsky}},\ }\href
  {https://doi.org/doi.org/10.15407/ujpe66.6.478} {\bibfield  {journal}
  {\bibinfo  {journal} {Ukrainian Journal of Physics}\ }\textbf {\bibinfo
  {volume} {66}},\ \bibinfo {pages} {478} (\bibinfo {year} {2021})}\BibitemShut
  {NoStop}%
\bibitem [{\citenamefont {Yi\u{g}it}\ and\ \citenamefont
  {Medvedev}(2015)}]{Yigit2015983}%
  \BibitemOpen
  \bibfield  {author} {\bibinfo {author} {\bibfnamefont {E.}~\bibnamefont
  {Yi\u{g}it}}\ and\ \bibinfo {author} {\bibfnamefont {A.~S.}\ \bibnamefont
  {Medvedev}},\ }\href
  {https://doi.org/https://doi.org/10.1016/j.asr.2014.11.020} {\bibfield
  {journal} {\bibinfo  {journal} {Advances in Space Research}\ }\textbf
  {\bibinfo {volume} {55}},\ \bibinfo {pages} {983} (\bibinfo {year}
  {2015})}\BibitemShut {NoStop}%
\bibitem [{\citenamefont {Kaladze}\ and\ \citenamefont
  {Misra}(2023)}]{Kaladze2023thermal}%
  \BibitemOpen
  \bibfield  {author} {\bibinfo {author} {\bibfnamefont {T.~D.}\ \bibnamefont
  {Kaladze}}\ and\ \bibinfo {author} {\bibfnamefont {A.~P.}\ \bibnamefont
  {Misra}},\ }\href {https://doi.org/10.1016/j.physleta.2023.128990} {\bibfield
   {journal} {\bibinfo  {journal} {Physics Letters A}\ }\textbf {\bibinfo
  {volume} {480}},\ \bibinfo {pages} {128990} (\bibinfo {year}
  {2023})}\BibitemShut {NoStop}%
\bibitem [{\citenamefont {Zhao}\ \emph {et~al.}(2024)\citenamefont {Zhao},
  \citenamefont {Wu}, \citenamefont {Duan}, \citenamefont {Wang}, \citenamefont
  {Li}, \citenamefont {Duan},\ and\ \citenamefont {Kang}}]{Zhao2024turbulence}%
  \BibitemOpen
  \bibfield  {author} {\bibinfo {author} {\bibfnamefont {Y.}~\bibnamefont
  {Zhao}}, \bibinfo {author} {\bibfnamefont {D.}~\bibnamefont {Wu}}, \bibinfo
  {author} {\bibfnamefont {L.}~\bibnamefont {Duan}}, \bibinfo {author}
  {\bibfnamefont {J.}~\bibnamefont {Wang}}, \bibinfo {author} {\bibfnamefont
  {J.}~\bibnamefont {Li}}, \bibinfo {author} {\bibfnamefont {L.}~\bibnamefont
  {Duan}},\ and\ \bibinfo {author} {\bibfnamefont {Q.}~\bibnamefont {Kang}},\
  }\href {https://doi.org/10.1063/5.0173929} {\bibfield  {journal} {\bibinfo
  {journal} {Physics of Fluids}\ }\textbf {\bibinfo {volume} {36}},\ \bibinfo
  {pages} {015132} (\bibinfo {year} {2024})}\BibitemShut {NoStop}%
\bibitem [{\citenamefont {Sharifi~Ghazijahani}\ and\ \citenamefont
  {Cierpka}(2024)}]{Sharifi2024spatiotemporal}%
  \BibitemOpen
  \bibfield  {author} {\bibinfo {author} {\bibfnamefont {M.}~\bibnamefont
  {Sharifi~Ghazijahani}}\ and\ \bibinfo {author} {\bibfnamefont
  {C.}~\bibnamefont {Cierpka}},\ }\href {https://doi.org/10.1063/5.0191403}
  {\bibfield  {journal} {\bibinfo  {journal} {Physics of Fluids}\ }\textbf
  {\bibinfo {volume} {36}},\ \bibinfo {pages} {035120} (\bibinfo {year}
  {2024})}\BibitemShut {NoStop}%
\bibitem [{\citenamefont {Kaladze}\ \emph {et~al.}(2022)\citenamefont
  {Kaladze}, \citenamefont {Misra}, \citenamefont {Roy},\ and\ \citenamefont
  {Chatterjee}}]{kaladze2022}%
  \BibitemOpen
  \bibfield  {author} {\bibinfo {author} {\bibfnamefont {T.~D.}\ \bibnamefont
  {Kaladze}}, \bibinfo {author} {\bibfnamefont {A.~P.}\ \bibnamefont {Misra}},
  \bibinfo {author} {\bibfnamefont {A.}~\bibnamefont {Roy}},\ and\ \bibinfo
  {author} {\bibfnamefont {D.}~\bibnamefont {Chatterjee}},\ }\href
  {https://doi.org/10.1016/j.asr.2022.02.014} {\bibfield  {journal} {\bibinfo
  {journal} {Advances in Space Research}\ }\textbf {\bibinfo {volume} {69}},\
  \bibinfo {pages} {3374} (\bibinfo {year} {2022})}\BibitemShut {NoStop}%
\bibitem [{\citenamefont {Shavit}\ \emph {et~al.}(2025)\citenamefont {Shavit},
  \citenamefont {B\"uhler},\ and\ \citenamefont {Shatah}}]{shavit2025}%
  \BibitemOpen
  \bibfield  {author} {\bibinfo {author} {\bibfnamefont {M.}~\bibnamefont
  {Shavit}}, \bibinfo {author} {\bibfnamefont {O.}~\bibnamefont {B\"uhler}},\
  and\ \bibinfo {author} {\bibfnamefont {J.}~\bibnamefont {Shatah}},\ }\href
  {https://doi.org/10.1103/PhysRevLett.134.054101} {\bibfield  {journal}
  {\bibinfo  {journal} {Phys. Rev. Lett.}\ }\textbf {\bibinfo {volume} {134}},\
  \bibinfo {pages} {054101} (\bibinfo {year} {2025})}\BibitemShut {NoStop}%
\bibitem [{\citenamefont {Vallis}(2017)}]{vallis2017}%
  \BibitemOpen
  \bibfield  {author} {\bibinfo {author} {\bibfnamefont {G.~K.}\ \bibnamefont
  {Vallis}},\ }\href {https://doi.org/10.1017/9781107588417} {\emph {\bibinfo
  {title} {Atmospheric and Oceanic Fluid Dynamics: Fundamentals and Large-Scale
  Circulation}}},\ \bibinfo {edition} {2nd}\ ed.\ (\bibinfo  {publisher}
  {Cambridge University Press},\ \bibinfo {year} {2017})\BibitemShut {NoStop}%
\bibitem [{\citenamefont {Shaikh}\ and\ \citenamefont
  {Shukla}(2009)}]{shaikh2009_simulation}%
  \BibitemOpen
  \bibfield  {author} {\bibinfo {author} {\bibfnamefont {D.}~\bibnamefont
  {Shaikh}}\ and\ \bibinfo {author} {\bibfnamefont {P.~K.}\ \bibnamefont
  {Shukla}},\ }\href {https://doi.org/10.1017/S0022377808007411} {\bibfield
  {journal} {\bibinfo  {journal} {Journal of Plasma Physics}\ }\textbf
  {\bibinfo {volume} {75}},\ \bibinfo {pages} {133–144} (\bibinfo {year}
  {2009})}\BibitemShut {NoStop}%
\bibitem [{\citenamefont {Shaikh}\ and\ \citenamefont
  {Zank}(2006)}]{Shaikh_2006}%
  \BibitemOpen
  \bibfield  {author} {\bibinfo {author} {\bibfnamefont {D.}~\bibnamefont
  {Shaikh}}\ and\ \bibinfo {author} {\bibfnamefont {G.~P.}\ \bibnamefont
  {Zank}},\ }\href {https://doi.org/10.1086/503833} {\bibfield  {journal}
  {\bibinfo  {journal} {The Astrophysical Journal}\ }\textbf {\bibinfo {volume}
  {640}},\ \bibinfo {pages} {L195} (\bibinfo {year} {2006})}\BibitemShut
  {NoStop}%
\bibitem [{\citenamefont {dong Ko}\ and\ \citenamefont {Kim}(2003)}]{ko2003}%
  \BibitemOpen
  \bibfield  {author} {\bibinfo {author} {\bibfnamefont {Y.}~\bibnamefont {dong
  Ko}}\ and\ \bibinfo {author} {\bibfnamefont {C.-B.}\ \bibnamefont {Kim}},\
  }in\ \href {https://doi.org/10.1109/PLASMA.2003.1228961} {\emph {\bibinfo
  {booktitle} {The 30th International Conference on Plasma Science, 2003. ICOPS
  2003. IEEE Conference Record - Abstracts.}}}\ (\bibinfo {year} {2003})\ p.\
  \bibinfo {pages} {351}\BibitemShut {NoStop}%
\bibitem [{\citenamefont {Kraichnan}(1967)}]{kraichnan1967}%
  \BibitemOpen
  \bibfield  {author} {\bibinfo {author} {\bibfnamefont {R.~H.}\ \bibnamefont
  {Kraichnan}},\ }\href {https://doi.org/10.1063/1.1762301} {\bibfield
  {journal} {\bibinfo  {journal} {Phys. Fluids}\ }\textbf {\bibinfo {volume}
  {10}},\ \bibinfo {pages} {1417} (\bibinfo {year} {1967})}\BibitemShut
  {NoStop}%
\bibitem [{\citenamefont {Nastrom}\ and\ \citenamefont
  {Gage}(1985)}]{nastrom1985}%
  \BibitemOpen
  \bibfield  {author} {\bibinfo {author} {\bibfnamefont {G.~D.}\ \bibnamefont
  {Nastrom}}\ and\ \bibinfo {author} {\bibfnamefont {K.~S.}\ \bibnamefont
  {Gage}},\ }\href
  {https://doi.org/10.1175/1520-0469(1985)042<0950:ACOAWS>2.0.CO;2} {\bibfield
  {journal} {\bibinfo  {journal} {J. Atmos. Sci.}\ }\textbf {\bibinfo {volume}
  {42}},\ \bibinfo {pages} {950} (\bibinfo {year} {1985})}\BibitemShut
  {NoStop}%
\bibitem [{USd()}]{USdata}%
  \BibitemOpen
  \href@noop {} {}\bibinfo {note} {Data is taken from the source ``U.S.
  Standard Atmosphere 1976" model. The temperature is 288.15 K at the sea level
  (0 km geopotential height) and the pressure is 101325 Pa. Details are
  available at
  \url{https://www.engineeringtoolbox.com/standard-atmosphere-d_604.html},
  1976.}\BibitemShut {Stop}%
\bibitem [{\citenamefont {Rogachevskii}\ and\ \citenamefont
  {Kleeorin}(2024)}]{Rogachevskii2024}%
  \BibitemOpen
  \bibfield  {author} {\bibinfo {author} {\bibfnamefont {I.}~\bibnamefont
  {Rogachevskii}}\ and\ \bibinfo {author} {\bibfnamefont {N.}~\bibnamefont
  {Kleeorin}},\ }\href {https://doi.org/10.1063/5.0188732} {\bibfield
  {journal} {\bibinfo  {journal} {Physics of Fluids}\ }\textbf {\bibinfo
  {volume} {36}},\ \bibinfo {pages} {026610} (\bibinfo {year}
  {2024})}\BibitemShut {NoStop}%
\end{thebibliography}%

\end{document}